\pdfoutput=1

\documentclass[12pt,a4paper]{article}

\usepackage{ifthen} 
\newboolean{pdflatex}
\setboolean{pdflatex}{true} 

\newboolean{articletitles}
\setboolean{articletitles}{true} 

\newboolean{uprightparticles}
\setboolean{uprightparticles}{false} 

\usepackage{microtype}
\usepackage{lineno}  
\usepackage{xspace} 

\usepackage{graphicx}  
\graphicspath{{./figs/}} 
\usepackage[abs]{overpic}

\usepackage[usenames,dvipsnames]{color}

\usepackage{amsmath} 
\usepackage{amssymb}
\usepackage{amsfonts}
\usepackage{upgreek} 

\newcommand*\patchAmsMathEnvironmentForLineno[1]{%
\expandafter\let\csname old#1\expandafter\endcsname\csname #1\endcsname
\expandafter\let\csname oldend#1\expandafter\endcsname\csname
end#1\endcsname
 \renewenvironment{#1}%
   {\linenomath\csname old#1\endcsname}%
   {\csname oldend#1\endcsname\endlinenomath}%
}
\newcommand*\patchBothAmsMathEnvironmentsForLineno[1]{%
  \patchAmsMathEnvironmentForLineno{#1}%
  \patchAmsMathEnvironmentForLineno{#1*}%
}
\AtBeginDocument{%
\patchBothAmsMathEnvironmentsForLineno{equation}%
\patchBothAmsMathEnvironmentsForLineno{align}%
\patchBothAmsMathEnvironmentsForLineno{flalign}%
\patchBothAmsMathEnvironmentsForLineno{alignat}%
\patchBothAmsMathEnvironmentsForLineno{gather}%
\patchBothAmsMathEnvironmentsForLineno{multline}%
}

\usepackage{hyperref}    
\usepackage[all]{hypcap} 

\def\lhcb {\mbox{LHCb}\xspace}
\def\ux85 {\mbox{UX85}\xspace}

\def\babar  {\mbox{BaBar}\xspace}

\ifthenelse{\boolean{uprightparticles}}%
{

 \def\Ppi         {\ensuremath{\uppi}\xspace}

 \def\PDelta      {\ensuremath{\Delta}\xspace}                 
 \def\PXi      {\ensuremath{\Xi}\xspace}                 
 \def\PLambda      {\ensuremath{\Lambda}\xspace}                 
 \def\PSigma      {\ensuremath{\Sigma}\xspace}                 
 \def\POmega      {\ensuremath{\Omega}\xspace}                 
 \def\PUpsilon      {\ensuremath{\Upsilon}\xspace}

 \def\PB      {\ensuremath{\mathrm{B}}\xspace}                 
                  
 \def\PD      {\ensuremath{\mathrm{D}}\xspace}

 \def\PK      {\ensuremath{\mathrm{K}}\xspace}

 \def\Pi      {\ensuremath{\mathrm{i}}\xspace}

}
{

 \def\Ppi         {\ensuremath{\pi}\xspace}

 \mathchardef\PDelta="7101
 \mathchardef\PXi="7104
 \mathchardef\PLambda="7103
 \mathchardef\PSigma="7106
 \mathchardef\POmega="710A
 \mathchardef\PUpsilon="7107
                  
 \def\PB      {\ensuremath{B}\xspace}                 
                  
 \def\PD      {\ensuremath{D}\xspace}

 \def\PK      {\ensuremath{K}\xspace}

 \def\Pi      {\ensuremath{i}\xspace}

}

\def\g      {\ensuremath{\Pgamma}\xspace}

\def\pion  {\ensuremath{\Ppi}\xspace}

\def\pip   {\ensuremath{\pion^+}\xspace}
\def\pim   {\ensuremath{\pion^-}\xspace}
\def\pipi  {\ensuremath{\pion^+\pion^-}\xspace}

\def\kaon  {\ensuremath{\PK}\xspace}
  \def\Kbar  {\kern 0.2em\overline{\kern -0.2em \PK}{}\xspace}

\def\Kz    {\ensuremath{\kaon^0}\xspace}
\def\Kzb   {\ensuremath{\Kbar^0}\xspace}
\def\KzKzb {\ensuremath{\Kz \kern -0.16em \Kzb}\xspace}
\def\Kp    {\ensuremath{\kaon^+}\xspace}
\def\Km    {\ensuremath{\kaon^-}\xspace}

\def\KpKm  {\ensuremath{\Kp \kern -0.16em \Km}\xspace}
\def\KS    {\ensuremath{\kaon^0_{\rm\scriptscriptstyle S}}\xspace}

  \def\Dbar    {\kern 0.2em\overline{\kern -0.2em \PD}{}\xspace}
\def\D       {\ensuremath{\PD}\xspace}

\def\Dz      {\ensuremath{\D^0}\xspace}
\def\Dzb     {\ensuremath{\Dbar^0}\xspace}
\def\DzDzb   {\ensuremath{\Dz {\kern -0.16em \Dzb}}\xspace}
\def\Dp      {\ensuremath{\D^+}\xspace}
\def\Dm      {\ensuremath{\D^-}\xspace}

\def\DpDm    {\ensuremath{\Dp {\kern -0.16em \Dm}}\xspace}

\def\B       {\ensuremath{\PB}\xspace}
\def\Bbar    {\ensuremath{\kern 0.18em\overline{\kern -0.18em \PB}{}}\xspace}

\def\Bu      {\ensuremath{\B^+}\xspace}
\def\Bub     {\ensuremath{\B^-}\xspace}
\def\Bp      {\ensuremath{\Bu}\xspace}
\def\Bm      {\ensuremath{\Bub}\xspace}
\def\Bpm     {\ensuremath{\B^\pm}\xspace}

  \def\Y#1S{\ensuremath{\PUpsilon{(#1S)}}\xspace}

\def\Lbar {\ensuremath{\kern 0.1em\overline{\kern -0.1em\PLambda}}\xspace}

\def\BF         {{\ensuremath{\cal B}\xspace}}

\def\BR         {\BF}
\newcommand{\decay}[2]{\ensuremath{#1\!\to #2}\xspace}         

\def\to                 {\ensuremath{\rightarrow}\xspace}

\def\CP                {\ensuremath{C\!P}\xspace}

\def\AT#1     {\ensuremath{A_{\mathrm{T}}^{#1}}\xspace}           

\def\C#1      {\ensuremath{\mathcal{C}_{#1}}\xspace}                       
\def\Cp#1     {\ensuremath{\mathcal{C}_{#1}^{'}}\xspace}                    
\def\Ceff#1   {\ensuremath{\mathcal{C}_{#1}^{\mathrm{(eff)}}}\xspace}        
\def\Cpeff#1  {\ensuremath{\mathcal{C}_{#1}^{'\mathrm{(eff)}}}\xspace}       
\def\Ope#1    {\ensuremath{\mathcal{O}_{#1}}\xspace}                       
\def\Opep#1   {\ensuremath{\mathcal{O}_{#1}^{'}}\xspace}                    

\newcommand{\tev}{\ensuremath{\mathrm{\,Te\kern -0.1em V}}\xspace}
\newcommand{\gev}{\ensuremath{\mathrm{\,Ge\kern -0.1em V}}\xspace}
\newcommand{\mev}{\ensuremath{\mathrm{\,Me\kern -0.1em V}}\xspace}
\newcommand{\kev}{\ensuremath{\mathrm{\,ke\kern -0.1em V}}\xspace}
\newcommand{\ev}{\ensuremath{\mathrm{\,e\kern -0.1em V}}\xspace}
\newcommand{\gevc}{\ensuremath{{\mathrm{\,Ge\kern -0.1em V\!/}c}}\xspace}
\newcommand{\mevc}{\ensuremath{{\mathrm{\,Me\kern -0.1em V\!/}c}}\xspace}
\newcommand{\gevcc}{\ensuremath{{\mathrm{\,Ge\kern -0.1em V\!/}c^2}}\xspace}
\newcommand{\gevgevcccc}{\ensuremath{{\mathrm{\,Ge\kern -0.1em V^2\!/}c^4}}\xspace}
\newcommand{\mevcc}{\ensuremath{{\mathrm{\,Me\kern -0.1em V\!/}c^2}}\xspace}

\def\invfb   {\ensuremath{\mbox{\,fb}^{-1}}\xspace}

\def\gsim{{~\raise.15em\hbox{$>$}\kern-.85em
          \lower.35em\hbox{$\sim$}~}\xspace}
\def\lsim{{~\raise.15em\hbox{$<$}\kern-.85em
          \lower.35em\hbox{$\sim$}~}\xspace}

\def\tell1  {TELL1\xspace}
\def\ukl1   {UKL1\xspace}

\usepackage{cite} 
\usepackage{mciteplus}

\newcommand{\figs}{figs}

\newcommand{\omcl}{\ensuremath{1-{\rm CL}}\xspace}

\newcommand{\calL}{\ensuremath{\mathcal{L}}\xspace}

\newcommand{\gDKCentral}	{\ensuremath{72.0}\xspace}
\newcommand{\gDKCentralPM}	{\ensuremath{\gDKCentral^{+14.7}_{-15.6}}\xspace}
\newcommand{\gDKCentralPMC}	{\ensuremath{\gDKCentralPM}\xspace}
\newcommand{\gDKOnesig} 	{\ensuremath{[   56.4,    86.7]}\xspace}
\newcommand{\gDKTwosig} 	{\ensuremath{[   42.6,    99.6]}\xspace}
\newcommand{\gDKOnesigC} 	{\ensuremath{\gDKOnesig}\xspace} 
\newcommand{\gDKTwosigC} 	{\ensuremath{\gDKTwosig}\xspace}
\newcommand{\rbDKCentral}  {\ensuremath{0.089}\xspace}
\newcommand{\rbDKOnesig}   {\ensuremath{[  0.080,   0.098]}\xspace}
\newcommand{\rbDKTwosig}   {\ensuremath{[  0.071,   0.107]}\xspace}
\newcommand{\dbDKCentral}  {\ensuremath{112}\xspace}
\newcommand{\dbDKOnesig}   {\ensuremath{[     96,     126]}\xspace}
\newcommand{\dbDKTwosig}   {\ensuremath{[     80,     136]}\xspace}

\newcommand{\gDpiCentralI}  {\ensuremath{18.9}\xspace}

\newcommand{\gDpiOnesigI}   {\ensuremath{[    8.9,    80.2]}\xspace}
\newcommand{\gDpiOnesigII}  {\ensuremath{[  169.1,   175.7]}\xspace}

\newcommand{\gDpiOnesigIC}	 {\ensuremath{[7.4, 99.2]}\xspace} 
\newcommand{\gDpiOnesigIIC}	 {\ensuremath{[167.9, 176.4]}\xspace}

\newcommand{\dbDpiCentralI} {\ensuremath{261}\xspace} 
\newcommand{\dbDpiOnesigI}  {\ensuremath{[    249,     331]}\xspace} 
\newcommand{\dbDpiOnesigII} {\ensuremath{[    213,     229]}\xspace} 

\newcommand{\rbDpiCentral } {\ensuremath{0.015}\xspace}
\newcommand{\rbDpiOnesig}   {\ensuremath{[  0.006,   0.056]}\xspace}
\newcommand{\rbDpiTwosig}   {\ensuremath{[  0.001,   0.073]}\xspace}

\newcommand{\gDKDpiCentral}    {\ensuremath{72.6}\xspace}

\newcommand{\gDKDpiOnesig}     {\ensuremath{[   56.7,    81.7]}\xspace}
\newcommand{\gDKDpiTwosig}     {\ensuremath{[   41.2,    92.3]}\xspace}
\newcommand{\gDKDpiOnesigC}		{\ensuremath{[55.4, 82.3]}\xspace} 
\newcommand{\gDKDpiTwosigC}	 	{\ensuremath{[40.2, 92.7]}\xspace}
\newcommand{\rbDKDpiCentral}   {\ensuremath{0.089}\xspace}
\newcommand{\rbDKDpiOnesig}    {\ensuremath{[  0.080,   0.097]}\xspace}
\newcommand{\rbDKDpiTwosig}    {\ensuremath{[  0.071,   0.105]}\xspace}
\newcommand{\dbDKDpiCentral}   {\ensuremath{112}\xspace}
\newcommand{\dbDKDpiOnesig}    {\ensuremath{[     96,     125]}\xspace}
\newcommand{\dbDKDpiTwosig}    {\ensuremath{[     79,     136]}\xspace}
\newcommand{\rbpiDKDpiCentral} {\ensuremath{0.015}\xspace}
\newcommand{\rbpiDKDpiOnesig}  {\ensuremath{[  0.006,   0.027]}\xspace}
\newcommand{\rbpiDKDpiTwosig}  {\ensuremath{[  0.002,   0.036]}\xspace}
\newcommand{\dbpiDKDpiCentral} {\ensuremath{315}\xspace} 
\newcommand{\dbpiDKDpiOnesig}  {\ensuremath{[    269,     332]}\xspace} 

\newcommand{\BmDzpi}  {\texorpdfstring{\decay{\Bm}{\Dz\pim}}			{B- -> D0pi-}}
\newcommand{\BmDzK}   {\texorpdfstring{\decay{\Bm}{\Dz\Km}}			{B- -> D0K-}}

\newcommand{\BmDzbpi} {\texorpdfstring{\decay{\Bm}{\Dzb\pim}}			{B- -> D0bar pi-}}
\newcommand{\BmDzbK}  {\texorpdfstring{\decay{\Bm}{\Dzb\Km}}			{B- -> D0bar K-}}

\newcommand{\BpDzh}   {\texorpdfstring{\decay{\Bp}{\Dz h^+}}			{B+ -> D0h+}}
\newcommand{\BpDzbh}  {\texorpdfstring{\decay{\Bp}{\Dzb h^+}}			{B+ -> D0barh+}}
\newcommand{\BpmDh}   {\texorpdfstring{\decay{\Bpm}{D h^\pm}}			{B+- -> Dh+-}}
\newcommand{\BpmDpi}  {\texorpdfstring{\decay{\Bpm}{D \pi^\pm}}		{B+- -> Dpi+-}}
\newcommand{\BpmDK}   {\texorpdfstring{\decay{\Bpm}{D K^\pm}}			{B+- -> DK+-}}

\newcommand{\DKpipipi} {\texorpdfstring{\ensuremath{D\to K\pi\pi\pi}}{D -> K3pi}\xspace}

\newcommand{\DzKpiFav} {\texorpdfstring{\ensuremath{\Dz\to\Km\pip}}{D0 -> K-pi+}\xspace}
\newcommand{\DzKpiSup} {\texorpdfstring{\ensuremath{\Dz\to\pim\Kp}}{D0 -> pi-K+}\xspace}

\newcommand{\DzKpppFav}{\texorpdfstring{\ensuremath{\Dz\to\Km\pip\pim\pip}}{D -> K3pi (fav)}\xspace}
\newcommand{\DzKpppSup}{\texorpdfstring{\ensuremath{\Dz\to\pim\Kp\pim\pip}}{D -> K3pi (sup)}\xspace}
\newcommand{\Dhh}      {\texorpdfstring{\ensuremath{D\to hh}}{D -> hh}\xspace}

\newcommand{\DzK}      {\texorpdfstring{\ensuremath{DK^\pm}}{DK}\xspace}
\newcommand{\Dzpi}     {\texorpdfstring{\ensuremath{D\pi^\pm}}{Dpi}\xspace}

\newcommand{\KK}   		{\texorpdfstring{\ensuremath{\Kp\Km}}{K+K-}\xspace}

\renewcommand{\g}{\texorpdfstring{\ensuremath{\gamma}}{gamma}\xspace}

\newcommand{\rbh}  {\texorpdfstring{\ensuremath{r_B^h}\xspace}{rBh}}
\newcommand{\rbhsq}{\texorpdfstring{\ensuremath{(r_B^{h})^2}\xspace}{rBh2}}
\newcommand{\dbh}  {\ensuremath{\delta_B^h}\xspace}

\newcommand{\rb}  {\texorpdfstring{\ensuremath{r_B^K}\xspace}{rBK}}
\newcommand{\rbsq}{\texorpdfstring{\ensuremath{(r_B^{K})^2}\xspace}{rBK2}}
\newcommand{\db}  {\ensuremath{\delta_B^K}\xspace}

\newcommand{\rbpi}  {\texorpdfstring{\ensuremath{r_B^{\pi}}\xspace}{rBpi}}
\newcommand{\rbpisq}{\texorpdfstring{\ensuremath{(r_B^{\pi})^2}\xspace}{rBpi2}}
\newcommand{\dbpi}  {\ensuremath{\delta_B^{\pi}}\xspace}

\newcommand{\rdFlv}  {\ensuremath{r_{f}}\xspace}
\newcommand{\rdFlvsq}{\ensuremath{r_{f}^2}\xspace}
\newcommand{\ddFlv}  {\ensuremath{\delta_{f}}\xspace}

\newcommand{\rdKpi}  {\ensuremath{r_{K\pi}}\xspace}

\newcommand{\ddKpi}  {\ensuremath{\delta_{K\pi}}\xspace}

\newcommand{\rdKppp}{\ensuremath{r_{K3\pi}}\xspace}

\newcommand{\ddKppp}{\ensuremath{\delta_{K3\pi}}\xspace}
\newcommand{\kdKppp}{\ensuremath{\kappa_{K3\pi}}\xspace}

\newcommand{\DAcpKK}  {\texorpdfstring{\ensuremath{A_{\CP}^{\rm dir}(KK)}\xspace}{AcpDir(KK)}}
\newcommand{\DAcpPipi}{\texorpdfstring{\ensuremath{A_{\CP}^{\rm dir}(\pi\pi)}\xspace}{AcpDir(pipi}}
\newcommand{\DAcphh}  {\texorpdfstring{\ensuremath{A_{\CP}^{\rm dir}}\xspace}{Acp(D->fCP)}}
\newcommand{\xd}{\texorpdfstring{\ensuremath{x_D}\xspace}{xD}}
\newcommand{\yd}{\texorpdfstring{\ensuremath{y_D}\xspace}{yD}}

\newcommand{\dRD}{\texorpdfstring{\ensuremath{R_D}\xspace}{RD}}
\newcommand{\xdprimesq}{\texorpdfstring{\ensuremath{x_D^{\prime2}}\xspace}{xD'2}}
\newcommand{\ydprime}{\texorpdfstring{\ensuremath{y'_D}\xspace}{yD'}}

\newcommand{\xpm}{\ensuremath{x_\pm}\xspace}

\newcommand{\ypm}{\ensuremath{y_\pm}\xspace}

\usepackage{longtable} 

\begin{document}

\renewcommand{\thefootnote}{\fnsymbol{footnote}}
\setcounter{footnote}{1}

\begin{titlepage}
\pagenumbering{roman}

\vspace*{-1.5cm}
\centerline{\large EUROPEAN ORGANISATION FOR NUCLEAR RESEARCH (CERN)}
\vspace*{1.5cm}
\hspace*{-0.5cm}
\begin{tabular*}{\linewidth}{lc@{\extracolsep{\fill}}r}
\ifthenelse{\boolean{pdflatex}}
{\vspace*{-2.7cm}\mbox{\!\!\!\includegraphics[width=.14\textwidth]{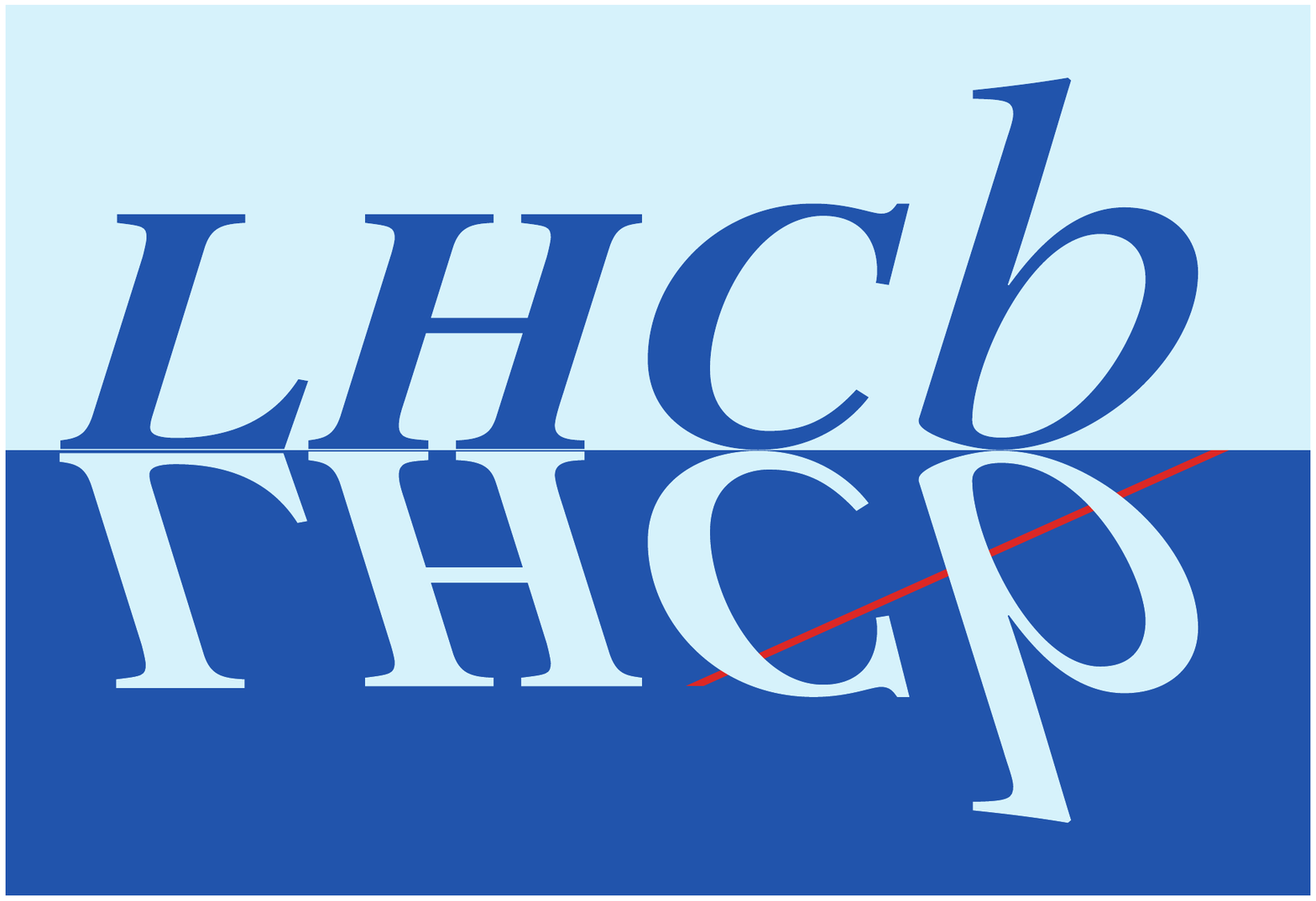}} & &}%
{\vspace*{-1.2cm}\mbox{\!\!\!\includegraphics[width=.12\textwidth]{figs/lhcb-logo.eps}} & &}%
\\
 & & CERN-PH-EP-2013-079 \\  
 & & LHCb-PAPER-2013-020 \\  
 & & 24. July 2013 \\ 
 & & \\
\end{tabular*}

\vspace*{0.25cm}

{\bf\boldmath\huge
\begin{center}
  Measurement of the CKM angle \g from a combination of \BpmDh analyses
\end{center}
}

\vspace{0.25cm}

\begin{center}
The LHCb collaboration\footnote{Authors are listed on the following pages.}
\end{center}

\vspace{0.25cm}

\begin{abstract}
A combination of three \lhcb measurements of
the CKM angle $\gamma$ is presented. The decays \BpmDK and \BpmDpi are used,
where $D$ denotes an admixture of \Dz and \Dzb mesons, 
decaying into $\Kp\Km$, \pipi, $K^\pm \pi^\mp$,
$K^\pm \pi^\mp \pi^\pm \pi^\mp$, $\KS\pi^+\pi^-$, or $\KS K^+K^-$ final states.
All measurements use a dataset corresponding to $1.0\invfb$ of integrated luminosity.
Combining results from \BpmDK decays alone 
a best-fit value of $\g = \gDKCentral^\circ$ is found, and  
confidence intervals are set
\begin{align}
\g \in \gDKOnesigC^\circ \quad &{\rm at\ 68\%\,CL}\,,\nonumber\\
\g \in \gDKTwosigC^\circ \quad &{\rm at\ 95\%\,CL}\,.\nonumber
\end{align}
The best-fit value of \g found from a combination
of results from \BpmDpi decays alone, 
is $\g = \gDpiCentralI^\circ$, and the
confidence intervals
\begin{align}
\g \in 
\gDpiOnesigIC^\circ \quad \cup \quad
\gDpiOnesigIIC^\circ \quad &{\rm at\ 68\%\,CL}\,\nonumber
\end{align}
are set, without constraint at $95\%$ CL.
The combination of results from
\BpmDK and \BpmDpi decays gives a
best-fit value of
$\g = \gDKDpiCentral^\circ$ and the confidence intervals
\begin{align}
\g \in \gDKDpiOnesigC^\circ  \quad &{\rm at\ 68\%\,CL}\,,\nonumber\\
\g \in \gDKDpiTwosigC^\circ  \quad &{\rm at\ 95\%\,CL}\,\nonumber
\end{align}
are set. All values are expressed modulo $180^\circ$, and are obtained
taking into account the effect of \Dz--\Dzb mixing.
\end{abstract}

\vspace{0.25cm}

\begin{center}
  To be submitted to Phys.~Lett.~B.
\end{center}

\vspace{0.25cm}

{\footnotesize 
\centerline{\copyright~CERN on behalf of the \lhcb collaboration, license \href{http://creativecommons.org/licenses/by/3.0/}{CC-BY-3.0}.}}
\vspace*{2mm}

\end{titlepage}

\newpage
\setcounter{page}{2}
\mbox{~}
\newpage

\centerline{\large\bf LHCb collaboration}
\begin{flushleft}
\small
R.~Aaij$^{40}$, 
C.~Abellan~Beteta$^{35,n}$, 
B.~Adeva$^{36}$, 
M.~Adinolfi$^{45}$, 
C.~Adrover$^{6}$, 
A.~Affolder$^{51}$, 
Z.~Ajaltouni$^{5}$, 
J.~Albrecht$^{9}$, 
F.~Alessio$^{37}$, 
M.~Alexander$^{50}$
% ...
\\
\bigskip

{\footnotesize \it
% ...
$ ^{5}$Clermont Universit\'{e}, Universit\'{e} Blaise Pascal, CNRS/IN2P3, LPC, Clermont-Ferrand, France\\
$ ^{6}$CPPM, Aix-Marseille Universit\'{e}, CNRS/IN2P3, Marseille, France\\
% ...
$ ^{9}$Fakult\"{a}t Physik, Technische Universit\"{a}t Dortmund, Dortmund, Germany\\
% ...
$ ^{35}$Universitat de Barcelona, Barcelona, Spain\\
$ ^{36}$Universidad de Santiago de Compostela, Santiago de Compostela, Spain\\
$ ^{37}$European Organization for Nuclear Research (CERN), Geneva, Switzerland\\
% ...
$ ^{40}$Nikhef National Institute for Subatomic Physics, Amsterdam, The Netherlands\\
% ...
$ ^{45}$H.H. Wills Physics Laboratory, University of Bristol, Bristol, United Kingdom\\
% ...
$ ^{50}$School of Physics and Astronomy, University of Glasgow, Glasgow, United Kingdom\\
$ ^{51}$Oliver Lodge Laboratory, University of Liverpool, Liverpool, United Kingdom\\
% ...
$ ^{n}$LIFAELS, La Salle, Universitat Ramon Llull, Barcelona, Spain\\
% ...
}
\end{flushleft}

\cleardoublepage

\renewcommand{\thefootnote}{\arabic{footnote}}
\setcounter{footnote}{0}

\pagestyle{plain} 
\setcounter{page}{1}
\pagenumbering{arabic}

\section{Introduction}
\label{sec:Introduction}

The angle \g is defined as 
$\g = \arg \left[ -V^{\phantom{*}}_{ud}V^*_{ub}/(V^{\phantom{*}}_{cd}V^*_{cb}) \right]$,
where $V_{ij}$ are the elements of the Cabibbo-Kobayashi-Maskawa (CKM) 
matrix~\cite{Cabibbo:1963yz,*Kobayashi:1973fv}.
It is one of the angles of the unitarity
triangle
and is to date the least well known angle of
this triangle. At the same time it is the only angle that can be 
measured entirely with decays that only involve tree diagrams, so
its measurement is largely unaffected by the theoretical uncertainty,
which is $\mathcal{O}(10^{-6})$~\cite{Zupan:2011mn}.
Both Belle and \babar
have recently published averages of 
their measurements, each following
a frequentist treatment.
Belle measures $\g = (68^{+15}_{-14})^\circ$~\cite{Trabelsi:2013uj},
and \babar measures $\g = (69^{+17}_{-16})^\circ$~\cite{Lees:2013fk}.
In this work a combination of LHCb measurements is presented.
World averages have been computed by the
CKMfitter and UTfit groups, who obtain
$\g = (66 \pm 12)^\circ$~\cite{Charles:2004jd}, and 
$\g = (70.8 \pm 7.8)^\circ$~\cite{Bona:2005vz},
using a frequentist and
Bayesian treatment, respectively. These averages are dominated by measurements
performed at the $B$ factories, and part of all LHCb measurements
combined in this work are already included.

When measuring \g in tree decays, an important
channel is the
\BpmDK mode, where the symbol $D$ denotes
an admixture of \Dz and \Dzb mesons. The $D$ meson is reconstructed in a final state 
accessible to both flavour states,
thus exploiting interference between the $b\to u\bar{c}s$ and $b\to c\bar{u}s$
amplitudes.
Throughout this Letter, charge conjugation applies,
unless stated otherwise.
The measurements are categorised
by the $D$ meson final state: \CP eigenstates (GLW~\cite{Gronau:1991dp,Gronau:1990ra}),
quasi-flavour-specific states (ADS~\cite{Atwood:1996ci,Atwood:2000ck}), 
and self-conjugate three-body
final states (GGSZ~\cite{Giri:2003ty}).
The small theoretical uncertainty 
in the measurement of \g is
obtained in these decays because all hadronic parameters are
determined from data. The amplitude ratio
$\rb = |A(\BmDzbK)/A(\BmDzK)|$, 
plays a crucial
role as the uncertainty on \g scales roughly as $1/\rb$. It is 
measured to be $\rb \approx 0.1$~\cite{Trabelsi:2013uj,Lees:2013fk}.

Besides the \BpmDK channel, the \BpmDpi decay also exhibits some
sensitivity to \g. The theoretical framework is fully analogous
to the \BpmDK case. However, the respective amplitude ratio \rbpi
is expected to be an order of magnitude smaller than \rb, limiting the sensitivity.
In this Letter, information from \BpmDpi
decays is included in the combined measurement of \g
for the first time.
The hadronic parameters describing the $D$ decays are
determined from data. To better constrain these parameters, 
measurements by CLEO are 
included~\cite{Lowery:2009id}, that themselves contain inputs
from the Heavy Flavour Averaging Group (HFAG).

It has been shown that the determination of \g
from \BpmDh decays, where $h=\pi,K$, is affected by \Dz--\Dzb 
mixing~\cite{Silva:1999bd,Atwood:2000ck,Grossman:2005rp,Bondar:2010qs,Matteo}.
It enters in two parts of the analysis: in the description of
the $B$ decays ({\it e.g.} through the amplitude
$B^+\to \Dz K^+ \to \Dzb K^+ \to f K^+$, where $f$ denotes the $D$ final state), 
and in the determination
of the hadronic parameters that describe the $D$ decay.
Since $D$ mixing is now well established, its effect
is included in this combination; the CLEO measurement~\cite{Lowery:2009id}
also takes it into account explicitly.
The effect of $D$ mixing on the GLW, ADS, and GGSZ analyses is
reviewed in Ref.~\cite{Matteo}: it mostly affects the ADS analysis of \BpmDpi decays,
due to the small expected value of \rbpi.
The ADS analysis of \BpmDK decays receives a shift of $|\Delta\g|\lesssim 1^{\circ}$~\cite{Matteo}.
The Dalitz-model independent GGSZ analysis of \BpmDK is affected to a negligible
extent~\cite{Bondar:2010qs,Matteo}, and the GLW analyses of \BpmDh are affected at most at order
of $\mathcal{O}(\rb\sqrt{\xd^2+\yd^2})$~\cite{Matteo}, where the mixing parameters \xd and \yd are 
at the level of $10^{-2}$. Here, a $D$ mixing measurement by LHCb~\cite{LHCb_Dmixing}
is included, to further constrain \xd and \yd.

The effect of possible \CP violation in
$D$ decays to the \pipi and \KK final states~\cite{Aaij:2013fk, LHCb-CONF-2013-003} has been 
discussed in Refs.~\cite{Martone:2012nj, Bhattacharya:2013fk, Wang:2012ie}.
This changes the interpretation of the observables of the GLW method,
which is included as described in Sect.~\ref{sec:input_glwads}.

In this combination, the strategy is to maximise 
a total likelihood built from the product of the 
probability density functions (PDFs) $f_i$ of  
experimental observables $\vec{A}_i$
\begin{equation}
	\label{eq:comblh}
	\calL(\vec{\alpha}) = \prod_i f_i(\vec{A}_i^{\rm obs} | \vec{\alpha})\,,
\end{equation}
where the $\vec{A}_i^{\rm obs}$ are the measured values of the observables,
and $\vec{\alpha}$ is the set of parameters.
The subscript $i$ denotes the contributing inputs, summarised in 
Sects.~\ref{sec:input_glwads}--\ref{sec:input_cleo}.
For most of the input measurements it is assumed that the observables follow 
a Gaussian distribution
\begin{equation}
	f_i \propto \exp\left( -\frac{1}{2} (\vec{A}_i(\vec{\alpha})-\vec{A}_{i}^{\rm obs})^T \, 
	V_i^{-1} \, (\vec{A}_i(\vec{\alpha})-\vec{A}_{i}^{\rm obs}) \right)\,,
\end{equation}
where $V_i$ is the experimental covariance matrix.
In this combined measurement the statistical uncertainties
dominate the resulting confidence intervals. Therefore it is assumed that
the systematic fluctuations are also Gaussian, so that
$V_i = V_i^{\rm stat} + V_i^{\rm syst}$. Since not all off-diagonal
entries of $V_i^{\rm syst}$ have been published, they are assumed to be
zero in the nominal result. An overall systematic uncertainty is estimated due
to this assumption. Any other correlations across the statistically
independent input measurements are neglected.
For one pair of variables (\kdKppp, \ddKppp, described in Sect.~\ref{sec:inputs})
that shows highly non-Gaussian behaviour, 
the experimental likelihood is taken into account.
Table~\ref{tab:pars} defines all free parameters in the global
fit. The amplitude ratios are defined as those of the suppressed 
processes divided by the favoured ones.
Confidence intervals on \g and the most important hadronic
parameters are set using a frequentist procedure. The statistical coverage of this
procedure is evaluated.

\begin{table}[h]
\centering
\caption{\small Free parameters used in the combined fit. The phase differences
\ddKpi and \ddKppp are defined in accordance with Refs.~\cite{Lees:2013fk,Trabelsi:2013uj,Lowery:2009id},
they are shifted by $180^{\circ}$ with respect to the HFAG.
Also, \g gains a sign for the conjugated modes,
$A(\BpDzh)/A(\BpDzbh) = \rbh e^{i(\dbh+\g)}$, with $h=K,\pi$.
\label{tab:pars}}
\begin{tabular}{llc}
\hline
Decay & Description & Parameter \\
\hline\\[-2.5ex]
\BpmDh                          & \CP-violating weak phase & $\gamma$ \\[0.5ex]
                                & $\Gamma(\BmDzK)/\Gamma(\BmDzpi)$ & $R_{\rm cab}$\\[0.5ex]
\BpmDpi                         & $A(\BmDzbpi)/A(\BmDzpi) = \rbpi e^{i(\dbpi-\g)}$ & \rbpi, \dbpi \\[0.5ex]
\BpmDK                          & $A(\BmDzbK)/A(\BmDzK) = \rb e^{i(\db-\g)}$ & \rb, \db \\[0.5ex]
\hline\\[-2.5ex]       
$\Dz\to K^\pm\pi^\mp$           & $A(\DzKpiSup)/A(\DzKpiFav) = \rdKpi e^{-i\ddKpi}$  & \rdKpi, $\ddKpi$ \\[0.5ex]
                                & Cabibbo-favoured rate & $\Gamma(D\to K\pi)$ \\[0.5ex]
$\Dz\to K^\pm\pi^\mp\pi^+\pi^-$ & amplitude ratio and effective strong phase diff. & \rdKppp, $\ddKppp$ \\[0.5ex]
                                & coherence factor  & \kdKppp \\[0.5ex]
                                & Cabibbo-favoured rate & $\Gamma(D\to K\pi\pi\pi)$ \\[0.5ex]
$\Dz\to K^+K^-$                 & direct \CP asymmetry & \DAcpKK \\[0.5ex]
$\Dz\to \pip\pim$               & direct \CP asymmetry & \DAcpPipi \\[0.5ex]
\Dz--\Dzb                       & mixing parameters & \xd, \yd \\[0.5ex]
\hline
\end{tabular}
\end{table}

\section{Input measurements}
\label{sec:inputs}

The LHCb collaboration has published three analyses relevant to this paper
based on the data corresponding to an integrated luminosity of
$1.0\invfb$ using $pp$ collisions at a centre-of-mass energy
of $7\tev$, recorded in 2011.
They are a GGSZ measurement of \BpmDK decays, where the $D$ meson
is reconstructed in the $D\to \KS\pip\pim$ and $D\to\KS\Kp\Km$
final states~\cite{Aaij:2012hu};
a GLW/ADS measurement of \BpmDK and \BpmDpi
decays, where the $D$ meson is reconstructed in charged two-body
final states~\cite{Aaij:2012kz};
and an ADS measurement of \BpmDK and \BpmDpi
decays, where the $D$ meson is reconstructed in charged
four-body final states~\cite{Aaij:2013aa}.
In addition, inputs from a combination of experimental data performed by the HFAG, 
to constrain the effect of direct \CP violation in $D$ decays~\cite{HFAG}, and
measurements from the LHCb collaboration~\cite{LHCb_Dmixing} and the
CLEO collaboration~\cite{Lowery:2009id}, to constrain the hadronic parameters 
of the $D$ system, are included. Ref.~\cite{Lowery:2009id} includes itself inputs by the HFAG.

\subsection{Measurements from \boldmath \texorpdfstring{$B^\pm\to D[\to \KS h^+h^-]K^\pm$}{B->D[->KShh]K} decays}
\label{sec:input_ggsz}

The GGSZ method~\cite{Giri:2003ty} proposes the use of self-conjugate 
three-body $D$ decays in the measurement of $\gamma$ from \BpmDK processes. 
The variables \xpm and \ypm,
defined as 
\begin{align}
x_\pm&=\rb\cos(\db\pm\gamma)\,,\\
y_\pm&=\rb\sin(\db\pm\gamma)\,,
\end{align}
are obtained from a fit to the Dalitz plane of $D\to\KS\pip\pim$
and $D\to\KS\Kp\Km$ decays, separately for \Bp and \Bm decays.
The measurement, performed by \lhcb, is reported in Ref.~\cite{Aaij:2012hu}.
The study makes no model-dependent assumption
on the variation of the strong phase of the $D\to\KS h^+h^-$ amplitudes, but 
instead uses measurements of this quantity from CLEO~\cite{Briere:2009uq}, as input.
The reported results are
\begin{align}
x_- &= (\phantom{-1}0.0\pm4.3\pm1.5\pm0.6)\times 10^{-2}\,,  \label{eq:ggsz1}\\
y_- &= (\phantom{-1}2.7\pm5.2\pm0.8\pm2.3)\times 10^{-2}\,,  \label{eq:ggsz2}\\
x_+ &= (         -10.3\pm4.5\pm1.8\pm1.4)\times 10^{-2}\,,   \label{eq:ggsz3}\\
y_+ &= (\phantom{1}{-}0.9\pm3.7\pm0.8\pm3.0)\times 10^{-2}\,,\label{eq:ggsz4}
\end{align}
where the first uncertainty is statistical, the second is systematic,
and the third is due to the external CLEO measurement.
The non-vanishing statistical correlations are $\rho(x_-,y_-) = -0.11$, $\rho(x_+,y_+) = +0.17$,
and the relevant systematic correlations are $\rho(x_-,y_-) = -0.05$, and $\rho(x_+,y_+) = +0.36$.

The GGSZ method can also be applied to \BpmDpi final states. In
Ref.~\cite{Aaij:2012hu} this was not performed, since these final states
were needed to control the efficiency variation across
the Dalitz plot. The effect of \Dz--\Dzb mixing in the measurement
of the $x_\pm$ and $y_\pm$ in Eqns.~\ref{eq:ggsz1}--\ref{eq:ggsz4}
is suppressed, leading to a negligible effect in the extraction of \g~\cite{Bondar:2010qs,Matteo}.

\subsection{Measurements from \boldmath \texorpdfstring{$B^\pm\to D[\to h^+h^-]h^\pm$}{B->D[->hh]h} decays}
\label{sec:input_glwads}

\newcommand{\fbar}{\ensuremath{\overline{f}}\xspace}

The $D$ decay modes considered in the analysis of two-body
$D$ final states~\cite{Aaij:2012kz}
are $D\to \Kp\Km$, $D\to\pip\pim$, the favoured decay
$D\to K^-\pi^+$, where the kaon charge matches that of the $h^\pm$ track
from the $B^\pm\to Dh^\pm$ decay (called $K\pi$ in the following), 
and the suppressed decay $D\to \pi^-K^+$, where the kaon charge is opposite that
of the $h^\pm$ track (called $\pi K$ in the following).
Building on the initial GLW/ADS ideas~\cite{Gronau:1991dp,Gronau:1990ra,Atwood:1996ci,Atwood:2000ck},
a set of 13 observables was defined by forming ratios of decay
rates, defined below, such that many systematic uncertainties cancel.
The charge-averaged ratios of \BpmDK and \BpmDpi decays are
\begin{align}
R_{K/\pi}^{f} &= \frac{\Gamma(\Bm\to D[\to f]K^-)   + \Gamma(\Bp\to D[\to \fbar]K^+)}
  	                  {\Gamma(\Bm\to D[\to f]\pi^-) + \Gamma(\Bp\to D[\to \fbar]\pi^+)}\,, \label{eq:rkpi}
\end{align}
where $f$ is the relevant final state. The ratios $R_{K/\pi}^{f}$
are related to \g and the hadronic parameters through
\begin{align}
R_{K/\pi}^{f} &= R_{\rm cab} 
                 \frac{1 + (\rb  \rdFlv)^2 + 2 \rb   \rdFlv \kappa\cos(\db   - \ddFlv) \cos\gamma + M^{K}_- + M^{K}_+}
                      {1 + (\rbpi\rdFlv)^2 + 2 \rbpi \rdFlv \kappa\cos(\dbpi - \ddFlv) \cos\gamma + M^{\pi}_- + M^{\pi}_+}\,, \label{eq:rkpi1}
\end{align}
for the favoured final state $f = K\pi$, 
where the coherence factor $\kappa$ in Eq.~\ref{eq:rkpi1} (and in all following
equations in this Section) is unity for two-body decays,
and through
\begin{align}
R_{K/\pi}^{f} &= R_{\rm cab} 
                 \frac{1 + \rbsq   + 2 \rb   \cos\db   \cos\gamma}
                      {1 + \rbpisq + 2 \rbpi \cos\dbpi \cos\gamma}\,, \label{eq:rkpi2}
\end{align}
for $f = KK, \pi\pi$. The $D$ mixing correction terms $M^h_{\pm}$ are,
at leading order in \xd and \yd, and neglecting \CP violation in $D$ mixing,
given by~\cite{Silva:1999bd}
\begin{align}
 M^h_{\pm} &= \left(\kappa\rdFlv(\rbhsq-1)\sin\ddFlv + \rbh(1-\rdFlvsq)\sin(\dbh\pm\g)\right)\, a_D \, \xd\nonumber\\
           &- \left(\kappa\rdFlv(\rbhsq+1)\cos\ddFlv + \rbh(1+\rdFlvsq)\cos(\dbh\pm\g)\right)\, a_D \, \yd\,.
\end{align}
The $D$ mixing corrections depend on the $D$ decay time acceptance
and resolution in the reconstruction of \BpmDh decays~\cite{Matteo}.
The coefficient $a_D$ parameterises their effect. It takes the value 
of $a_D=1$ in case of an ideal,
flat acceptance and negligible time resolution. For a 
realistic acceptance and resolution model
present in the GLW/ADS analysis of Ref.~\cite{Aaij:2012kz}, 
it is estimated to be $a_D=1.20\pm0.04$, where the uncertainty
can be safely neglected in this combination.
For \CP even final states of the $D$ meson, the mixing corrections cancel
exactly in Eq.~\ref{eq:rkpi2} (and~\ref{eq:acp2}), as in this case $\kappa=1$, $\rdFlv=1$, $\ddFlv=0$.
The charge asymmetries are
\begin{align}
	A_{h}^{f} &= \frac{\Gamma(\Bm\to D[\to f]h^-) - \Gamma(\Bp\to D[\to \fbar]h^+)}
	                  {\Gamma(\Bm\to D[\to f]h^-) + \Gamma(\Bp\to D[\to \fbar]h^+)} \,, 
\end{align}
which are related to \g and the hadronic parameters through
\begin{align}
    A_{h}^{f} &= \frac{2 \rbh \rdFlv \kappa\sin(\dbh - \ddFlv) \sin\gamma + M^{h}_- - M^{h}_+}
                      {1 + (\rbh\rdFlv)^2 + 2 \rbh \rdFlv \kappa\cos(\dbh - \ddFlv) \cos\gamma + M^{h}_- + M^{h}_+}\,, \label{eq:acp1}
\end{align}
for the favoured final state $f = K\pi$, and through
\begin{align}
 A_{h}^{f} &= \frac{2 \rbh \sin\dbh \sin\gamma}
                         {1 + \rbhsq + 2 \rbh \cos\dbh \cos\gamma}\,, \label{eq:acp2}
\end{align}
for $f = KK$, $\pi\pi$,
where \rbh denotes \rb and \rbpi.
Finally, the non charge-averaged ratios of suppressed and 
favoured $D$ final states are
\begin{align}
R_{h}^{\pm}  &= \frac{\Gamma(\Bpm\to D[\to f_{\rm sup}]h^\pm)}
	                 {\Gamma(\Bpm\to D[\to f]h^\pm)} \nonumber\\
	         &= \frac{\rdFlvsq + \rbhsq + 2 \rbh \rdFlv\kappa\cos(\dbh+\ddFlv\pm\gamma) - [M^{h}_{\pm}]_{\rm sup}}
	                 {1 + (\rbh\rdFlv)^2 + 2 \rbh \rdFlv\kappa\cos(\dbh-\ddFlv\pm\gamma) + M^{h}_{\pm}}\,,\label{eq:rpm}
\end{align}
where $f_{\rm sup}=\pi K$ is the suppressed final state,
and $f = K\pi$ the allowed one.
The suppressed $D$ mixing correction terms are given,
at leading order in \xd and \yd, by
\begin{align}
  [M^{h}_{\pm}]_{\rm sup}&= \left(\kappa\rdFlv(\rbhsq-1)\sin\ddFlv + \rbh(1-\rdFlvsq)\sin(\dbh\pm\g)\right)\, a_D \, \xd\nonumber\\
                         & +\left(\kappa\rdFlv(\rbhsq+1)\cos\ddFlv + \rbh(1+\rdFlvsq)\cos(\dbh\pm\g)\right)\, a_D \, \yd\,.
\end{align}
The combination makes use of all \g-sensitive observables 
determined in the GLW/ADS analysis. The full set, taken 
from the two-body analysis~\cite{Aaij:2012kz}, is
\begin{align*}
R_{K/\pi}^{K\pi}  &= \phantom{-} 0.0774 \phantom{0} \pm 0.0012  \phantom{0} \pm 0.0018\,, \\
R_{K/\pi}^{KK}    &= \phantom{-} 0.0773 \phantom{0} \pm 0.0030  \phantom{0} \pm 0.0018\,, \\
R_{K/\pi}^{\pi\pi}&= \phantom{-} 0.0803 \phantom{0} \pm 0.0056  \phantom{0} \pm 0.0017\,, \\
A_{\pi}^{K\pi}    &=            -0.0001 \phantom{0} \pm 0.0036  \phantom{0} \pm 0.0095\,, \\
A_{K}^{K\pi}      &= \phantom{-} 0.0044 \phantom{0} \pm 0.0144  \phantom{0} \pm 0.0174\,, \\
A_{K}^{KK}        &= \phantom{-} 0.148  \phantom{00}\pm 0.037   \phantom{00}\pm 0.010\,,\\
A_{K}^{\pi\pi}    &= \phantom{-} 0.135  \phantom{00}\pm 0.066   \phantom{00}\pm 0.010\,, \\
A_{\pi}^{KK}      &=            -0.020  \phantom{00}\pm 0.009   \phantom{00}\pm 0.012\,,\\
A_{\pi}^{\pi\pi}  &=            -0.001  \phantom{00}\pm 0.017   \phantom{00}\pm 0.010\,,\\
R_{K}^-           &= \phantom{-} 0.0073 \phantom{0} \pm 0.0023  \phantom{0} \pm 0.0004\,, \\
R_{K}^+           &= \phantom{-} 0.0232 \phantom{0} \pm 0.0034  \phantom{0} \pm 0.0007\,, \\
R_{\pi}^-         &= \phantom{-} 0.00469            \pm 0.00038             \pm 0.00008\,, \\
R_{\pi}^+         &= \phantom{-} 0.00352            \pm 0.00033             \pm 0.00007\,,
\end{align*}
where the first uncertainty is statistical and the second systematic.
Their statistical correlations, not previously published, are given
in Table~\ref{tab:glwadshhCor}.

\begin{table}[!b]
\scriptsize
\centering
\caption{\small Statistical correlations of the $\Bpm\to Dh^\pm$, 
\Dhh analysis~\cite{Aaij:2012kz}.}
\label{tab:glwadshhCor}
\begin{tabular}{lccccccccccccc}
\hline\\[-2.5ex]
     &  $A_{K}^{KK}$ & $A_{\pi}^{KK}$ &     $A_{K}^{\pi\pi}$ &   $A_{\pi}^{\pi\pi}$ &       $A_{K}^{K\pi}$ &     $A_{\pi}^{K\pi}$ & $R_{K/\pi}^{\pi\pi}$ &     $R_{K/\pi}^{KK}$ &   $R_{K/\pi}^{K\pi}$ &            $R_{K}^-$ &          $R_{\pi}^-$ &            $R_{K}^+$ &          $R_{\pi}^+$ \\
\hline\\[-2.5ex]
        $A_{K}^{KK}$ &       1 & -0.029 &      0 &      0 &      0 &      0 & -0.002 & -0.034 & -0.010 & -0.001 &      0 &      0 &      0 \\[1.0ex]
      $A_{\pi}^{KK}$ &         &      1 &      0 &      0 &      0 &      0 &      0 & -0.003 &      0 &      0 &      0 &      0 &      0 \\[1.0ex]
    $A_{K}^{\pi\pi}$ &         &        &      1 & -0.032 &      0 &      0 & -0.032 & -0.002 & -0.004 & -0.001 &      0 &      0 &      0 \\[1.0ex]
  $A_{\pi}^{\pi\pi}$ &         &        &        &      1 &      0 &      0 & -0.004 &      0 &      0 &      0 &      0 &      0 &      0 \\[1.0ex]
      $A_{K}^{K\pi}$ &         &        &        &        &      1 & -0.045 &      0 &      0 &  0.003 &  0.004 &      0 & -0.004 & -0.001 \\[1.0ex]
    $A_{\pi}^{K\pi}$ &         &        &        &        &        &      1 &      0 &      0 & -0.001 &  0.004 &  0.002 & -0.004 & -0.002 \\[1.0ex]
$R_{K/\pi}^{\pi\pi}$ &         &        &        &        &        &        &      1 &  0.013 &  0.029 &  0.003 &  0.003 &  0.001 &  0.003 \\[1.0ex]
    $R_{K/\pi}^{KK}$ &         &        &        &        &        &        &        &      1 &  0.053 &  0.005 &  0.005 &  0.002 &  0.004 \\[1.0ex]
  $R_{K/\pi}^{K\pi}$ &         &        &        &        &        &        &        &        &      1 & -0.038 &  0.016 & -0.093 &  0.014 \\[1.0ex]
           $R_{K}^-$ &         &        &        &        &        &        &        &        &        &      1 & -0.023 &  0.012 &  0.006 \\[1.0ex]
         $R_{\pi}^-$ &         &        &        &        &        &        &        &        &        &        &      1 &  0.005 &  0.008 \\[1.0ex]
           $R_{K}^+$ &         &        &        &        &        &        &        &        &        &        &        &      1 & -0.036 \\[1.0ex]
         $R_{\pi}^+$ &         &        &        &        &        &        &        &        &        &        &        &        &      1 \\[1.0ex]            
\hline
\end{tabular}
\end{table}

Direct \CP asymmetries in 
$\Dz\to \Kp\Km$ and $\Dz\to \pip\pim$ decays have been 
measured~\cite{Aaij:2013fk, LHCb-CONF-2013-003}.
While the effect on the charge averaged
ratios $R_{K/\pi}^{KK}$ and $R_{K/\pi}^{\pi\pi}$ is 
negligible~\cite{Bhattacharya:2013fk},
the observables $A_h^{KK}$
and $A_h^{\pi\pi}$ are modified by adding the respective direct \CP asymmetry
\DAcphh to the right-hand side of Eq.~\ref{eq:acp2}.
This is valid up to neglecting a small weak phase in the $D$ decay~\cite{Bhattacharya:2013fk}.
The HFAG results on \DAcphh~\cite{HFAG} are included in this combination
\begin{align*}
\DAcpKK &= (-0.31 \pm 0.24)\times 10^{-2}\,,\\
\DAcpPipi &= (+0.36 \pm 0.25)\times 10^{-2}\,.
\end{align*}
These quantities are correlated,
$\rho(\DAcpKK,\DAcpPipi) = +0.80$, and therefore
they are constrained to their observed values
by means of a two-dimensional correlated Gaussian PDF.
The inclusion of the result on $\DAcpKK-\DAcpPipi$~\cite{Aaij:2013fk},
which is statistically independent from the HFAG average,
is found to
have no effect on the combination.

\subsection{Measurements from \boldmath \texorpdfstring{$B^\pm\to D[\to K^\pm\pi^\mp\pi^+\pi^-]h^\pm$}{B->D[->K3pi]h} decays}
\label{sec:input_adsk3pi}

The $D$ four-body decay modes considered in the analysis of Ref.~\cite{Aaij:2013aa}
are the favoured $D\to K^- \pi^+\pi^-\pi^+$,
and the suppressed $D\to \pi^- K^+\pi^-\pi^+$ final states.
In a similar manner to the two-body GLW/ADS analysis, seven observables are defined
as ratios of decay rates. Their relations to \g and the hadronic
parameters are fully analogous and given by Eqs.~\ref{eq:rkpi1},~\ref{eq:acp1},
and~\ref{eq:rpm}, with $f = K\pi\pi\pi$ and 
$f_{\rm sup} = \pi K\pi\pi$. The \CP-violating effects are 
diluted due to the $D$ decay proceeding through a range of resonances
that can only interfere in limited regions of the four-body phase space. 
This dilution is accounted for by multiplying each interference term by a coherence factor $\kappa=\kdKppp$.
The $D$ decay time acceptance and resolution model is identical to that
present in the two-body GLW/ADS analysis of Ref.~\cite{Aaij:2012kz}.
The seven observables, taken from the four-body analysis reported in Ref.~\cite{Aaij:2013aa}, are
\begin{align*}
R_{K/\pi}^{K3\pi} & =  \phantom{-}0.0765 \phantom{0} \pm 0.0017  \phantom{0} \pm 0.0026\,, \\
A_{\pi}^{K3\pi}   & =            -0.006  \phantom{00}\pm 0.005   \phantom{00}\pm 0.010\,, \\
A_{K}^{K3\pi}     & =            -0.026  \phantom{00}\pm 0.020   \phantom{00}\pm 0.018\,, \\
R_{K-}^{K3\pi}    & =  \phantom{-}0.0071 \phantom{0} \pm 0.0034  \phantom{0} \pm 0.0008\,, \\
R_{K+}^{K3\pi}    & =  \phantom{-}0.0155 \phantom{0} \pm 0.0042  \phantom{0} \pm 0.0010\,, \\
R_{\pi-}^{K3\pi}  & =  \phantom{-}0.00400            \pm 0.00052             \pm 0.00011\,, \\
R_{\pi+}^{K3\pi}  & =  \phantom{-}0.00316            \pm 0.00046             \pm 0.00011\,,
\end{align*}
where the first uncertainty is statistical and the second systematic.
The statistical correlations between these variables, not previously published,
are presented in Table~\ref{tab:glwadsk3piCor}.

\begin{table}[!hbtp]
\scriptsize
\centering
\caption{\small Statistical correlations of the $\Bpm\to Dh^\pm$, \DKpipipi analysis~\cite{Aaij:2013aa}.}
\label{tab:glwadsk3piCor}
\begin{tabular}{lccccccc}
\hline\\[-2.5ex]
 & $R_{K/\pi}^{K3\pi}$  & $A_{K}^{K3\pi}$  & $A_{\pi}^{K3\pi}$  & $R_{K-}^{K3\pi}$  & $R_{K+}^{K3\pi}$  & $R_{\pi-}^{K3\pi}$  & $R_{\pi+}^{K3\pi}$ \\
\hline\\[-2.5ex]
$R_{K/\pi}^{K3\pi}$ &       1 &  0.003 &  0.001 & -0.060 & -0.024 &  0.017 &  0.021 \\[1.0ex]
    $A_{K}^{K3\pi}$ &         &      1 & -0.035 & -0.007 &  0.006 & -0.002 &  0.002 \\[1.0ex]
  $A_{\pi}^{K3\pi}$ &         &        &      1 & -0.006 &  0.008 & -0.002 &  0.005 \\[1.0ex]
  $R_{K-}^{K3\pi}$ &         &        &        &      1 &  0.043 &  0.006 &  0.029  \\[1.0ex]
  $R_{K+}^{K3\pi}$ &         &        &        &        &      1 &  0.022 &  0.025  \\[1.0ex]
$R_{\pi-}^{K3\pi}$ &         &        &        &        &        &      1 &  0.032  \\[1.0ex]
$R_{\pi+}^{K3\pi}$ &         &        &        &        &        &        &      1  \\[1.0ex]
\hline
\end{tabular}
\end{table}

\subsection{\boldmath Measurement of the hadronic parameters of the 
$D$ system from \texorpdfstring{$\Dz\to K^\pm\pi^\mp$}{D -> Kpi},
\texorpdfstring{$K^\pm\pi^\mp\pip\pim$}{D -> K3pi} decays by CLEO}
\label{sec:input_cleo}

The two- and four-body ADS measurements both reach their best sensitivity 
when combined with knowledge of the hadronic parameters of the $D$ decay. 
These are, for the $\Dz\to K^\pm\pi^\mp$ decays, the amplitude ratio \rdKpi and the strong 
phase difference \ddKpi. 
The hadronic parameters of the $\Dz\to K^\pm\pi^\mp\pip\pim$ decays are the ratio \rdKppp, the 
phase \ddKppp and the coherence factor\footnote{Note that Ref.~\cite{Lowery:2009id}
uses the symbol $R_{K3\pi}$ to denote the coherence factor.} \kdKppp. 
All of these parameters are constrained by a CLEO measurement~\cite{Lowery:2009id}, 
where a combined fit is performed, which includes information on the $D$ mixing 
parameters and the Cabibbo-favoured branching fractions of the $D$ decay 
through the following relationship
\begin{align}
\frac{\Gamma(\Dz\to f_{\rm sup})}{\Gamma(\Dz\to f_{\rm fav})} 
= \rdFlvsq \left[ 1 - \frac{y_D}{\rdFlv} \kappa \cos\ddFlv 
+ \frac{x_D}{\rdFlv} \kappa \sin\ddFlv + \frac{x_D^2+y_D^2}{2 \rdFlvsq} \right]\,,
\end{align}
where $\rdFlv=\rdKpi$ $(\rdKppp)$, $\ddFlv=\ddKpi$ $(\ddKppp)$, and
$\kappa=1$ $(\kdKppp)$, for $\Dz\to K^\pm\pi^\mp$ ($K^\pm\pi^\mp\pip\pim$) decays.
All of these parameters are included in the combination, although the 
dependence of \g on the $D$ mixing parameters and the Cabibbo-favoured 
branching fractions is small compared to the current statistical precision.
The central values and the uncertainties given in Table~\ref{tab:cleoobs} 
are reproduced from the analysis by the CLEO collaboration reported in 
Ref.~\cite{Lowery:2009id}.
The covariance matrix (see Table~VI in Ref.~\cite{Lowery:2009id}) is also used, 
though it is not reproduced here.
The parameters $(\ddKppp, \kdKppp)$ exhibit a non-Gaussian two-dimensional
likelihood (see Fig.~2b in Ref.~\cite{Lowery:2009id}), 
and this likelihood is used in the combination~\cite{cleoprivate}.
Their central values and profile-likelihood uncertainties are
$\kdKppp = 0.33^{+0.26}_{-0.23}$ and
$\ddKppp = (114^{+26}_{-23})^\circ$.
Correlations of \ddKppp and \kdKppp to other parameters are neglected.

\begin{table}[!htb]
\centering
\caption{\small Results of the CLEO measurement~\cite{Lowery:2009id}.}
\label{tab:cleoobs}
\begin{tabular}{lc}
\hline
Observable & Central value and uncertainty \\
\hline\\[-2.5ex]
\ddKpi            & $(-151.5^{+9.6}_{-9.5})^\circ$  \\
\xd               & $(0.96 \pm 0.25)\times 10^{-2}$ \\
\yd               & $(0.81 \pm 0.16)\times 10^{-2}$ \\
$\BR(\DzKpiFav)$  & $(3.89 \pm 0.05)\times 10^{-2}$ \\
$\BR(\DzKpiSup)$  & $(1.47 \pm 0.07)\times 10^{-4}$ \\
$\BR(\DzKpppFav)$ & $(7.96 \pm 0.19)\times 10^{-2}$ \\
$\BR(\DzKpppSup)$ & $(2.65 \pm 0.19)\times 10^{-4}$ \\
\hline
\end{tabular}                                    
\end{table}

\subsection{\boldmath Measurement from \texorpdfstring{$\Dz\to K^\pm\pi^\mp$}{D -> Kpi}
decays by LHCb}
\label{sec:input_lhcbdmixing}

The $D$ mixing parameters \xd and \yd are constrained in addition by
an LHCb measurement of $\Dz\to K^\pm\pi^\mp$ decays~\cite{LHCb_Dmixing}.
Three observables are defined, \dRD, \ydprime, and \xdprimesq, that
are related to the $D$ system parameters through the following
relationships
\begin{align}
        \dRD       &= \rdKpi^2\,, \\
        \ydprime   &= \xd\sin\ddKpi-\yd\cos\ddKpi\,, \\ 
        \xdprimesq &= \left(\xd\cos\ddKpi+\yd\sin\ddKpi\right)^2\,,
\end{align}
where a phase shift of $180^\circ$ was introduced to \ddKpi to be in accordance
with the phase convention adopted in this Letter.
In Ref.~\cite{LHCb_Dmixing}, the measured central values of the
observables are
$\dRD = (3.52 \pm 0.15)\times 10^{-3}$,
$\ydprime = (7.2 \pm 2.4)\times 10^{-3}$, and 
$\xdprimesq = (-0.09 \pm 0.13)\times 10^{-3}$,
where the error includes both statistical and systematic uncertainties.
These observables are strongly correlated,
$\rho(\dRD,\ydprime)=-0.95$,
$\rho(\ydprime,\xdprimesq)=-0.97$, and
$\rho(\xdprimesq,\dRD)=+0.88$.
They are included by means of a three-dimensional correlated
Gaussian PDF.

\section{Statistical interpretation}
\label{sec:statinterpretation}

The evaluation of this combination follows a frequentist approach. A $\chi^2$-function is defined as
$\chi^2(\vec{\alpha}) = -2 \ln \mathcal{L}(\vec{\alpha})$,
where $\mathcal{L}(\vec{\alpha})$ is defined in Eq.~\ref{eq:comblh}.
The best-fit point is given by the global minimum of the 
$\chi^2$-function, $\chi^2(\vec{\alpha}_{\min})$.
To evaluate the confidence level for a given value of a certain parameter,
say $\g=\g_0$ in the following,
the value of the $\chi^2$-function
at the new minimum is considered, $\chi^2(\vec{\alpha}'_{\min}(\g_0))$.
This also defines the profile
likelihood function $\hat{\calL}(\g_0) = \exp(-\chi^2(\vec{\alpha}'_{\min})/2)$.
Then a test statistic is defined as 
\mbox{$\Delta\chi^2 = \chi^2(\vec{\alpha}'_{\min}) - \chi^2(\vec{\alpha}_{\min})$}.
The $p$-value, or \omcl, is calculated by means of a Monte Carlo 
procedure, described in Ref.~\cite{woodroofe} 
and briefly recapitulated here.
For each value of $\g_0$:
\begin{enumerate}
  \item $\Delta\chi^2$ is calculated;
  \item a set of pseudoexperiments $\vec{A}_j$ is generated using
  Eq.~\ref{eq:comblh} with parameters $\vec{\alpha}$ set to $\vec{\alpha}'_{\min}$
  as the PDF;
  \item $\Delta\chi^{2\prime}$ of the pseudoexperiment is calculated 
  by replacing $\vec{A}_{\rm obs} \to \vec{A}_j$ and
  minimising with respect to $\vec{\alpha}$,
  once with \g as a free parameter, and once with \g fixed to $\gamma_0$;
  \item \omcl is calculated as the fraction of pseudoexperiments which
  perform worse ($\Delta\chi^2 < \Delta\chi^{2\prime}$) than the measured data.
\end{enumerate}
This method is sometimes known as the ``$\hat\mu$'', or the ``plug-in'' 
method. Its coverage cannot be guaranteed~\cite{woodroofe}
for the full parameter space, but is verified for the best-fit point. 
The reason is, that at each point $\gamma_0$, the nuisance parameters, 
{\em i.e.} the components of $\vec{\alpha}$ other than the parameter of interest,
are set to their best-fit values for this point, as opposed to 
computing an $n$-dimensional confidence belt, which is computationally very demanding.

In case of the CLEO 
likelihood for \kdKppp and \ddKppp, it is assumed that the true PDF,
for any assumed true value of \kdKppp and \ddKppp, can be described 
by a shifted version of the likelihood profile. 
In the non-physical range, $\kdKppp\notin [0,1]$, the likelihood profile
is not available. It is extrapolated into the non-physical range
using Gaussian tails that correspond
to the published uncertainties of the central value.
If $H(x,y)$ 
denotes the provided likelihood profile, with a maximum at position $(\hat{x}, \hat{y})$,
it is transformed as $f_i(x_{\rm obs},y_{\rm obs}|x,y) \propto 
H_i( x - x_{\rm obs} + \hat{x}, ~y - y_{\rm obs} + \hat{y})$,
with the abbreviation $(x,y)=(\kdKppp,\ddKppp)$.

\section{Results}
\label{sec:results}

\begin{figure}[!htb]
\centering
\begin{overpic}[width=.49\textwidth]{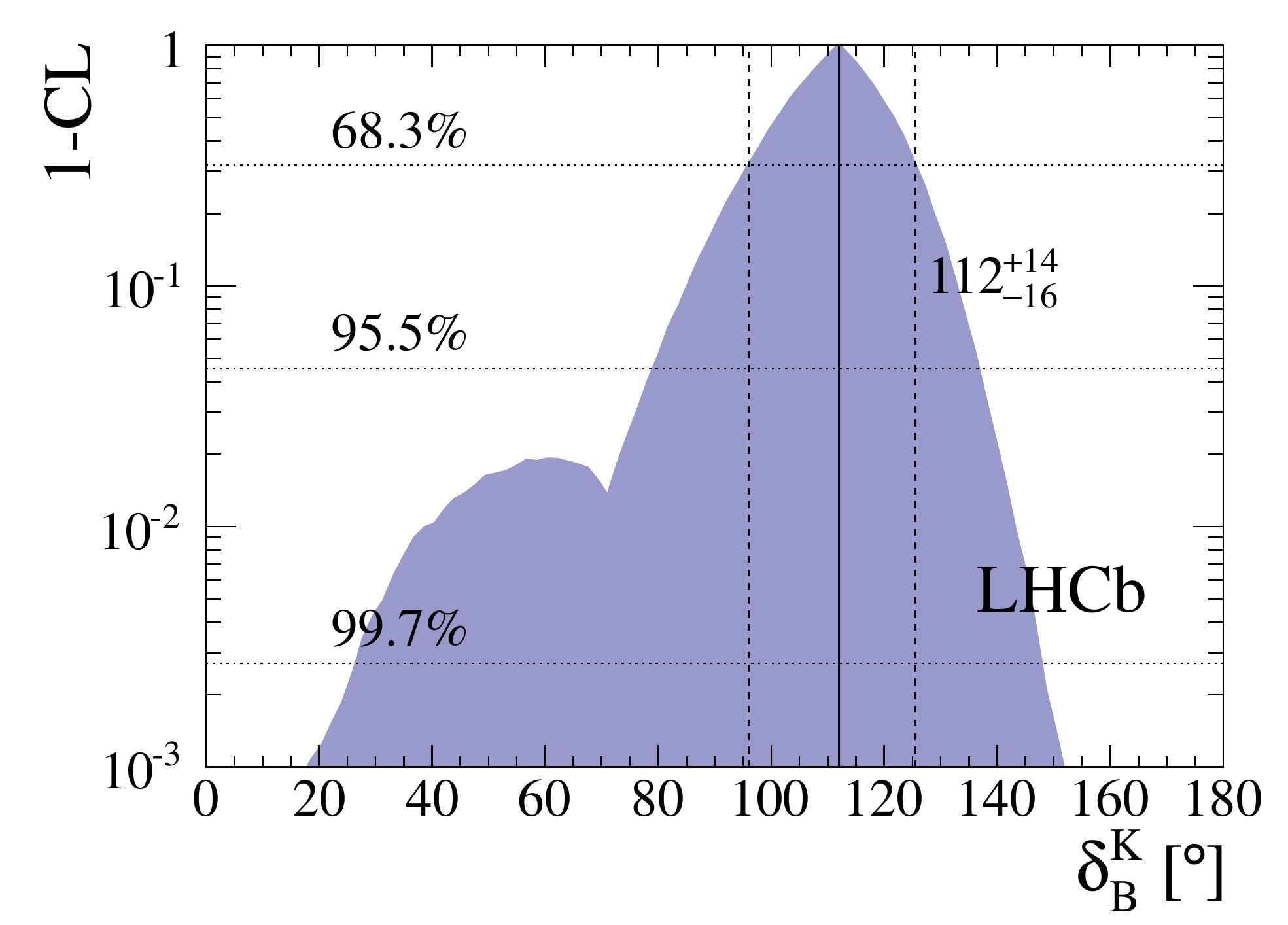}\put(190,135){a)}\end{overpic}
\begin{overpic}[width=.49\textwidth]{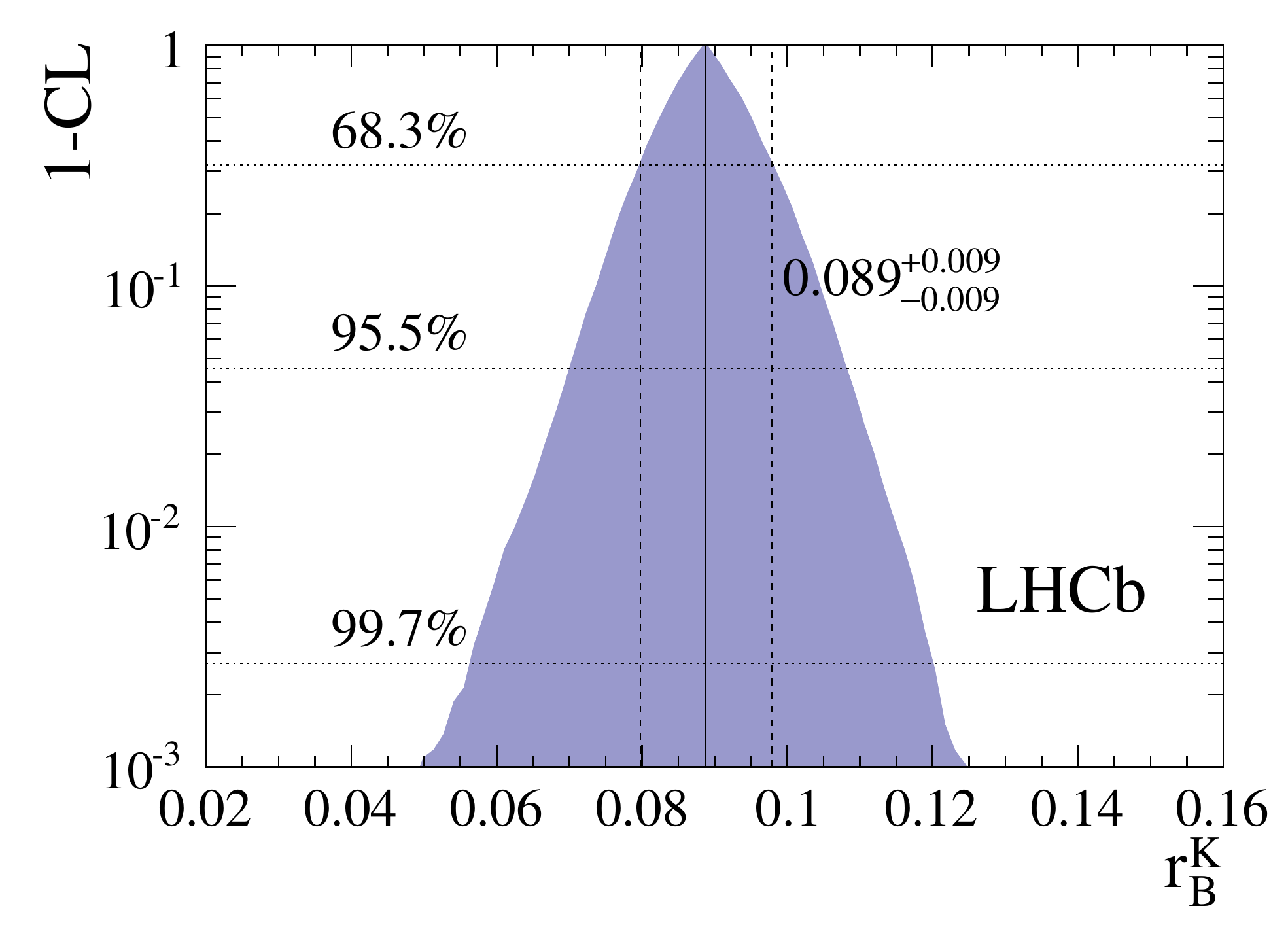}\put(190,135){b)}\end{overpic}\\
\begin{overpic}[width=.49\textwidth]{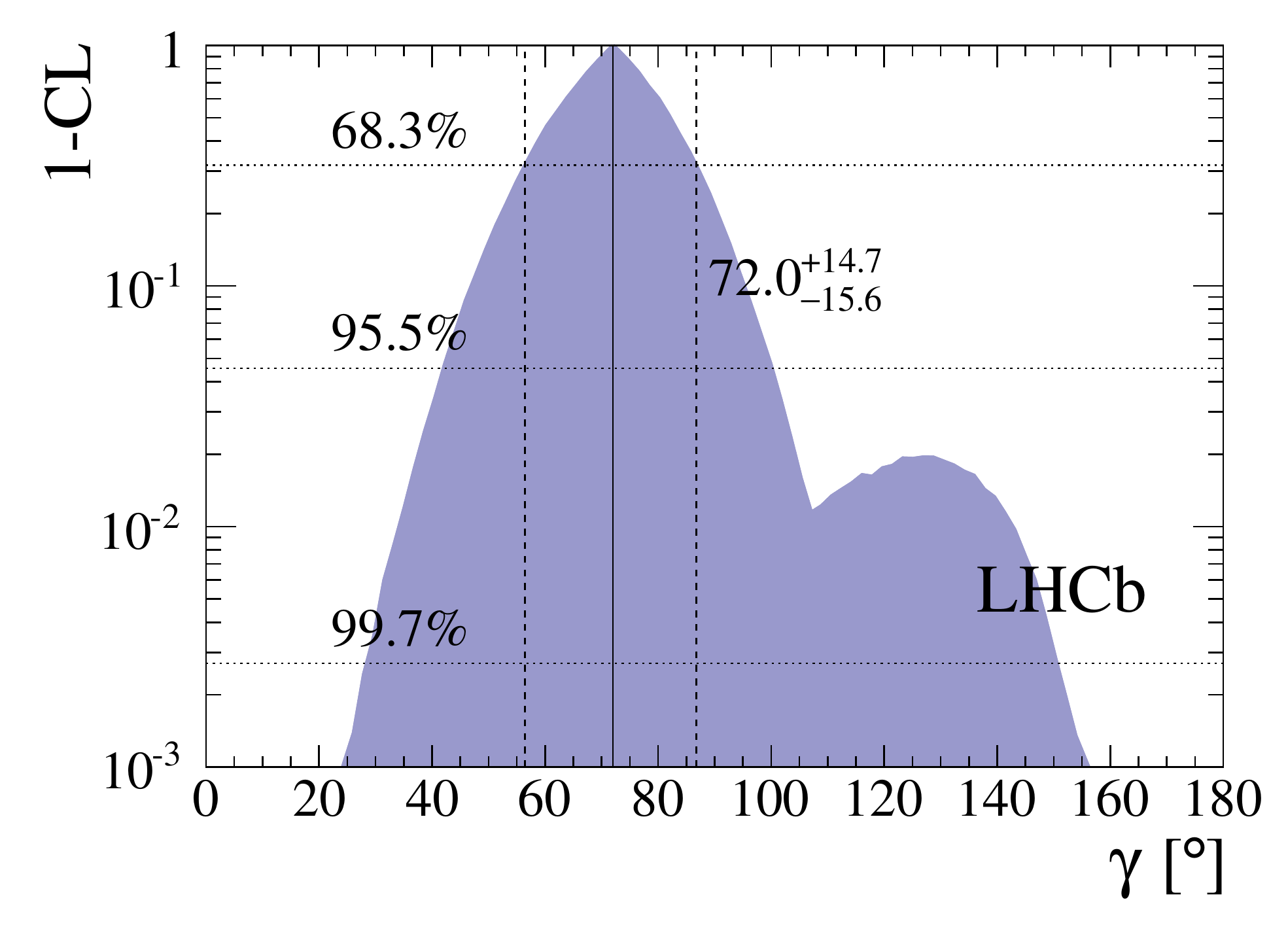}   \put(190,135){c)}\end{overpic}
\caption{\small Graphs showing \omcl for (a) \db, (b) \rb, and (c) \g, for the \DzK 
combination of the two- and four-body GLW/ADS and the \DzK GGSZ measurements.
The reported numbers correspond to the best-fit values and the uncertainties
are computed using the respective $68.3\%$ CL confidence interval shown
in Table~\ref{tab:dk}.}
\label{fig:dk1log}
\end{figure}

\begin{figure}[!htb]
\centering
\begin{overpic}[width=.49\textwidth]{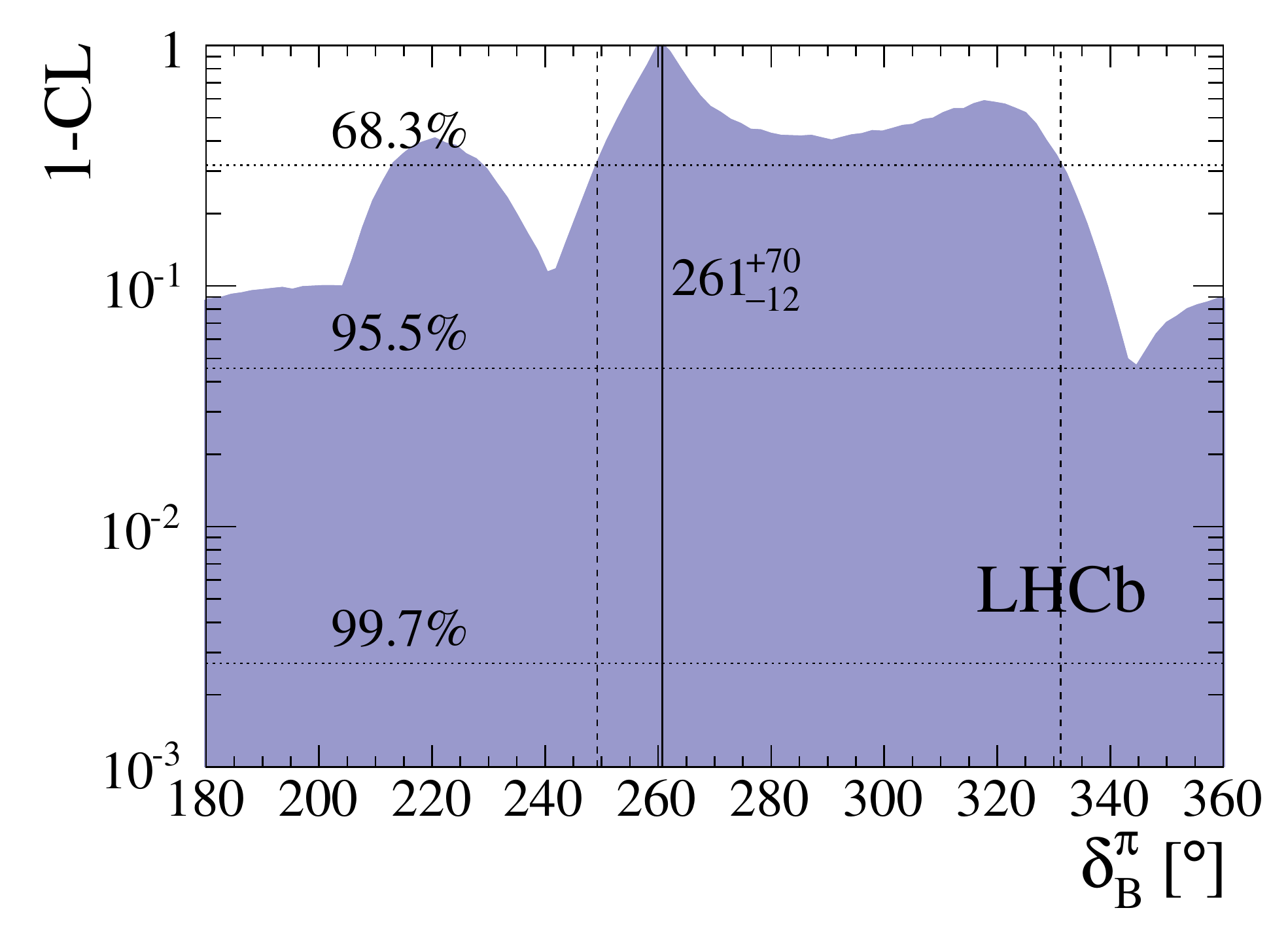}\put(190,135){a)}\end{overpic}
\begin{overpic}[width=.49\textwidth]{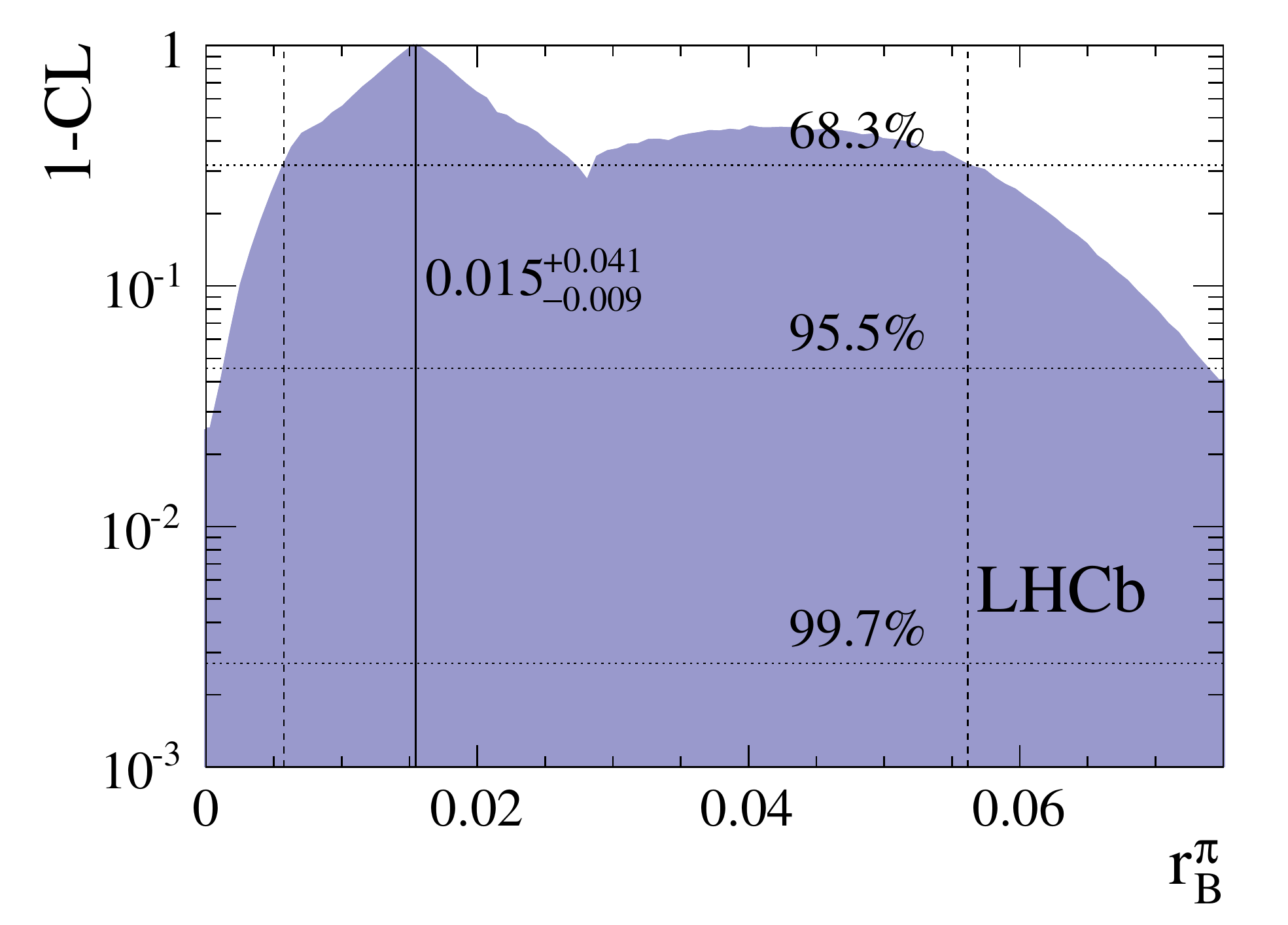}\put(190,135){b)}\end{overpic}\\
\begin{overpic}[width=.49\textwidth]{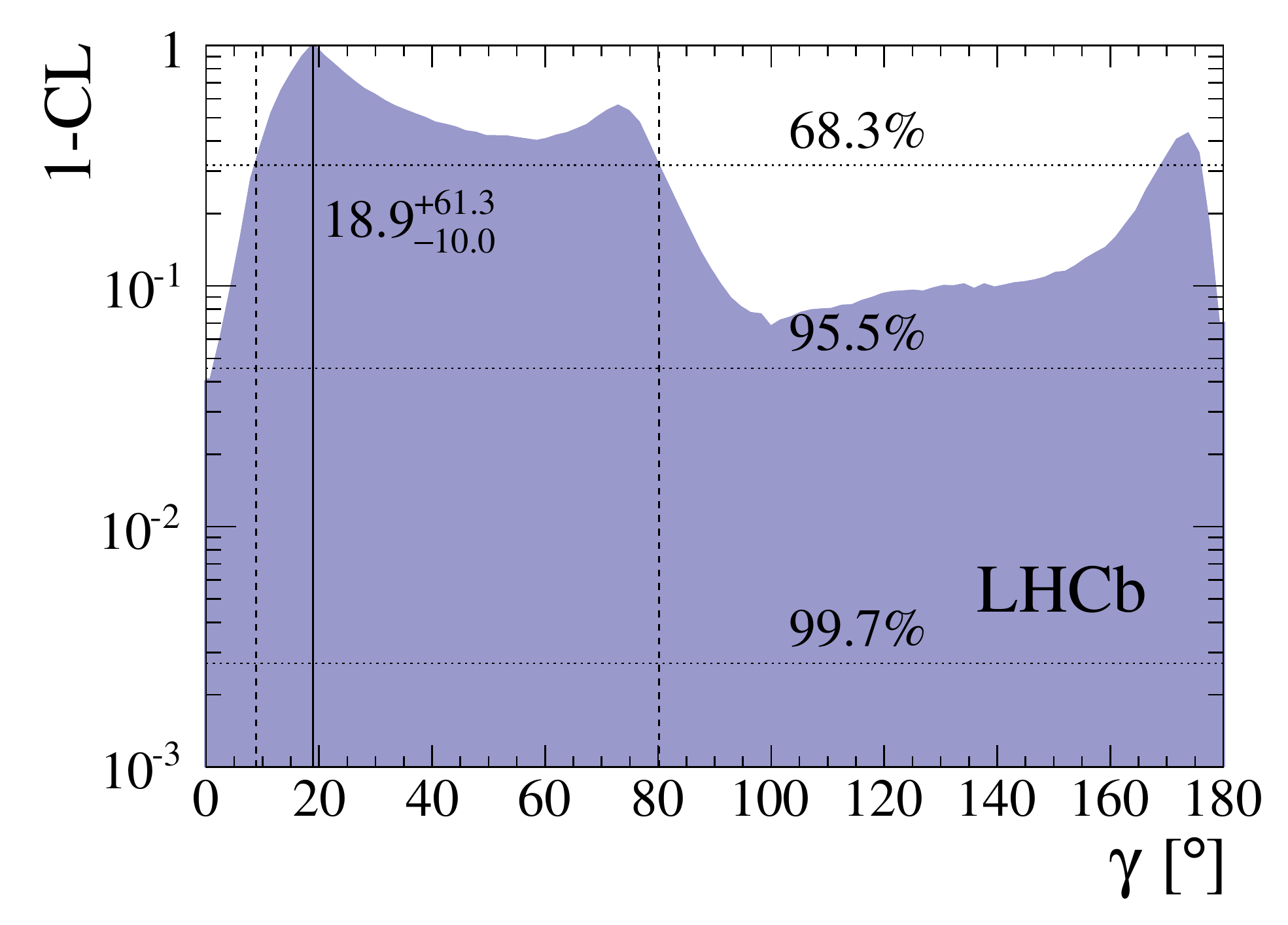}    \put(190,135){c)}\end{overpic}
\caption{\small Graphs showing \omcl for (a) \dbpi, (b) \rbpi, and (c) \g, for the \Dzpi 
combination of the two- and four-body GLW/ADS measurements.
The reported numbers correspond to the best-fit values and the uncertainties
are computed using appropriate $68.3\%$ CL confidence intervals shown
in Table~\ref{tab:dpi}.}
\label{fig:dpi1log}
\end{figure}

\begin{figure}[!htb]
\centering
\begin{overpic}[width=.49\textwidth]{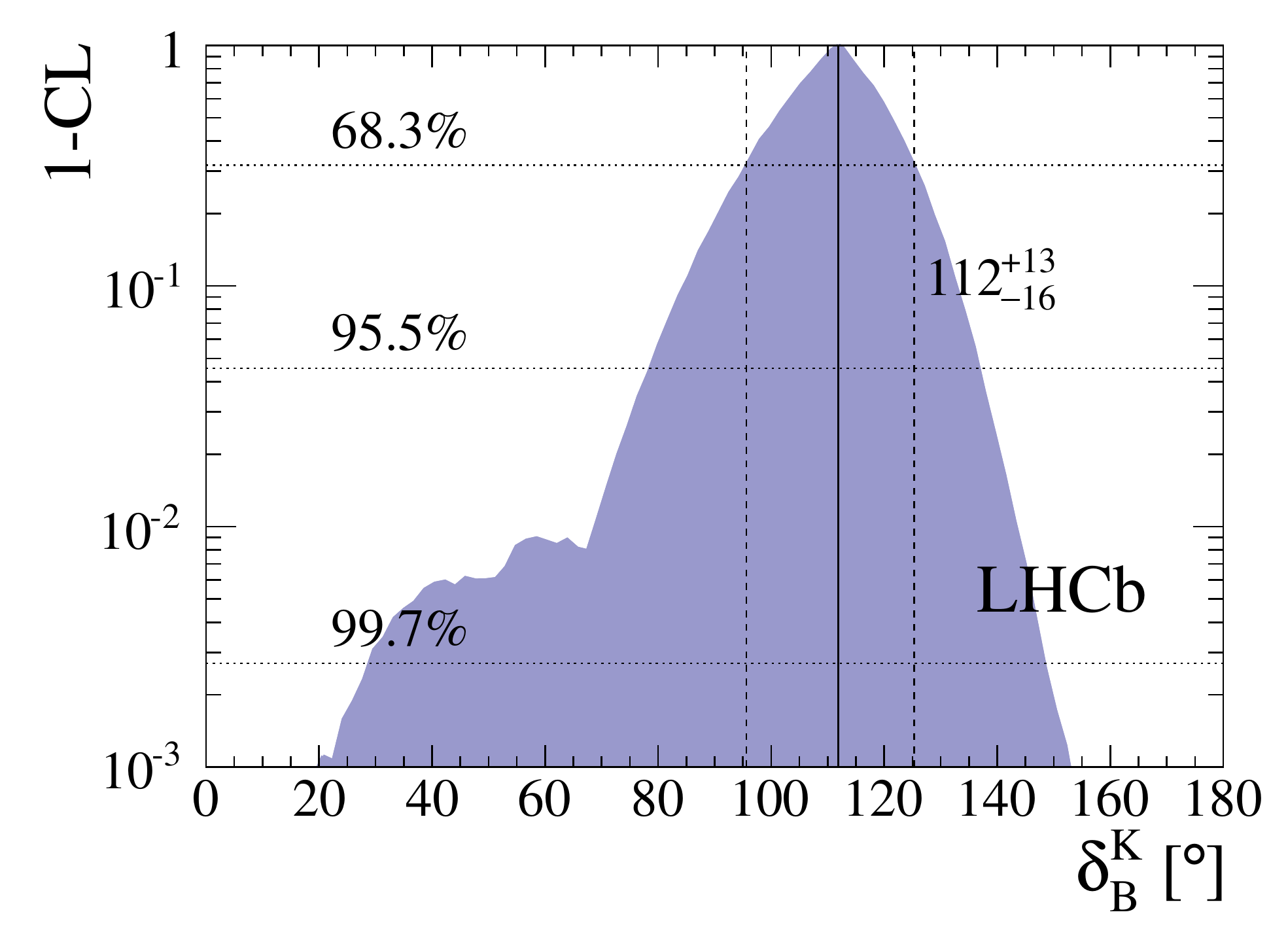} \put(190,135){a)}\end{overpic}
\begin{overpic}[width=.49\textwidth]{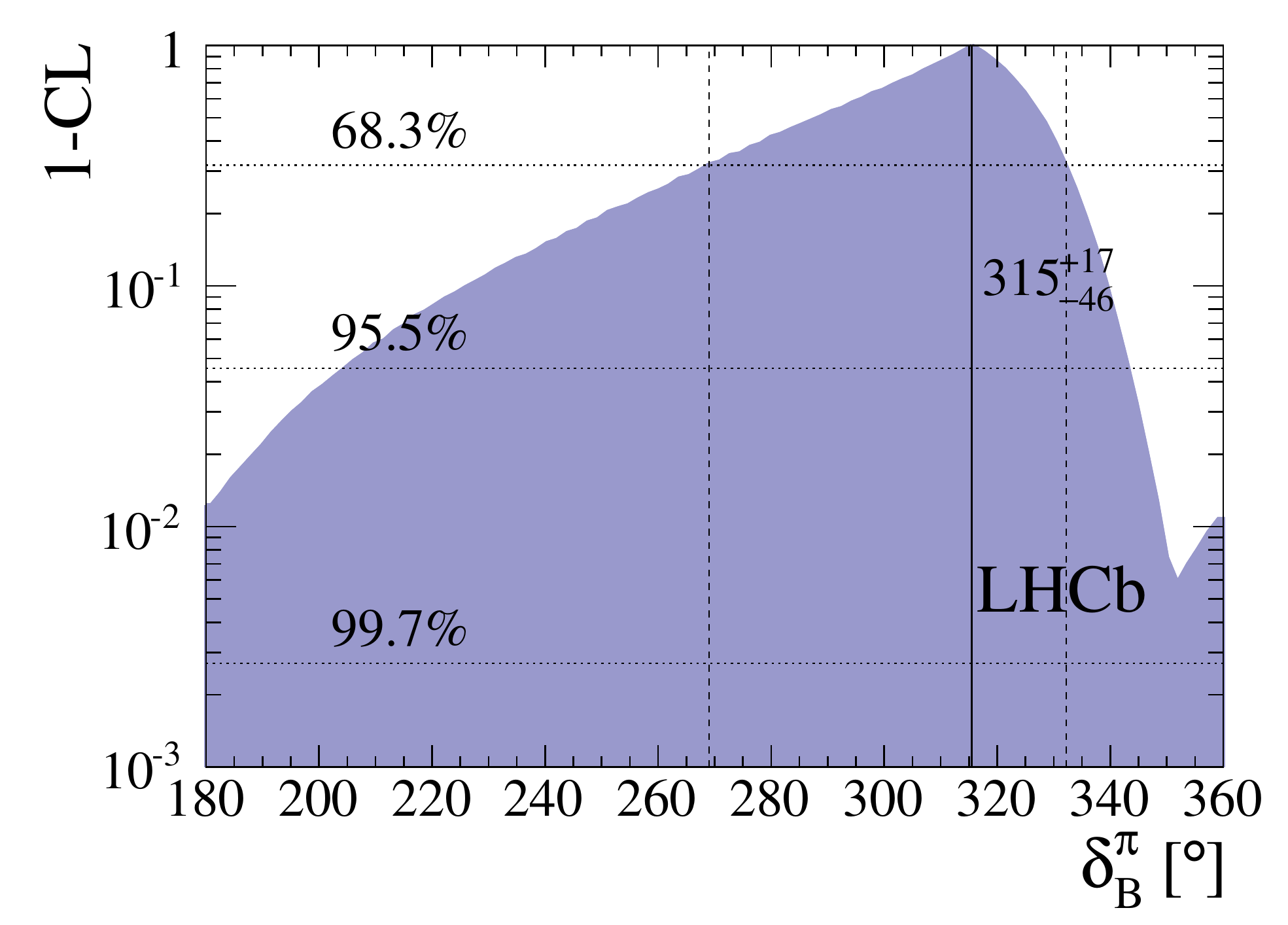}\put(190,135){b)}\end{overpic}\\
\begin{overpic}[width=.49\textwidth]{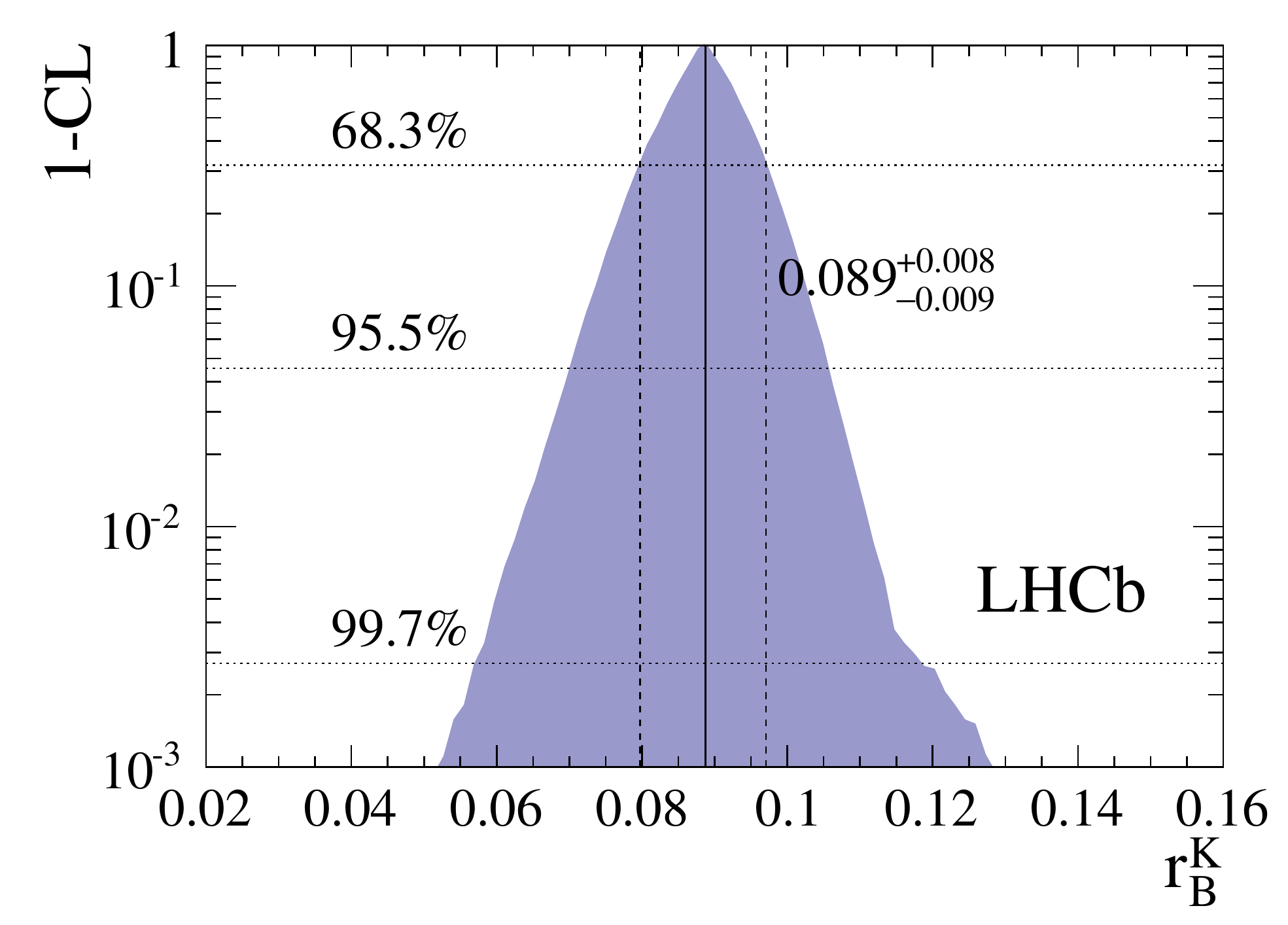} \put(190,135){c)}\end{overpic}
\begin{overpic}[width=.49\textwidth]{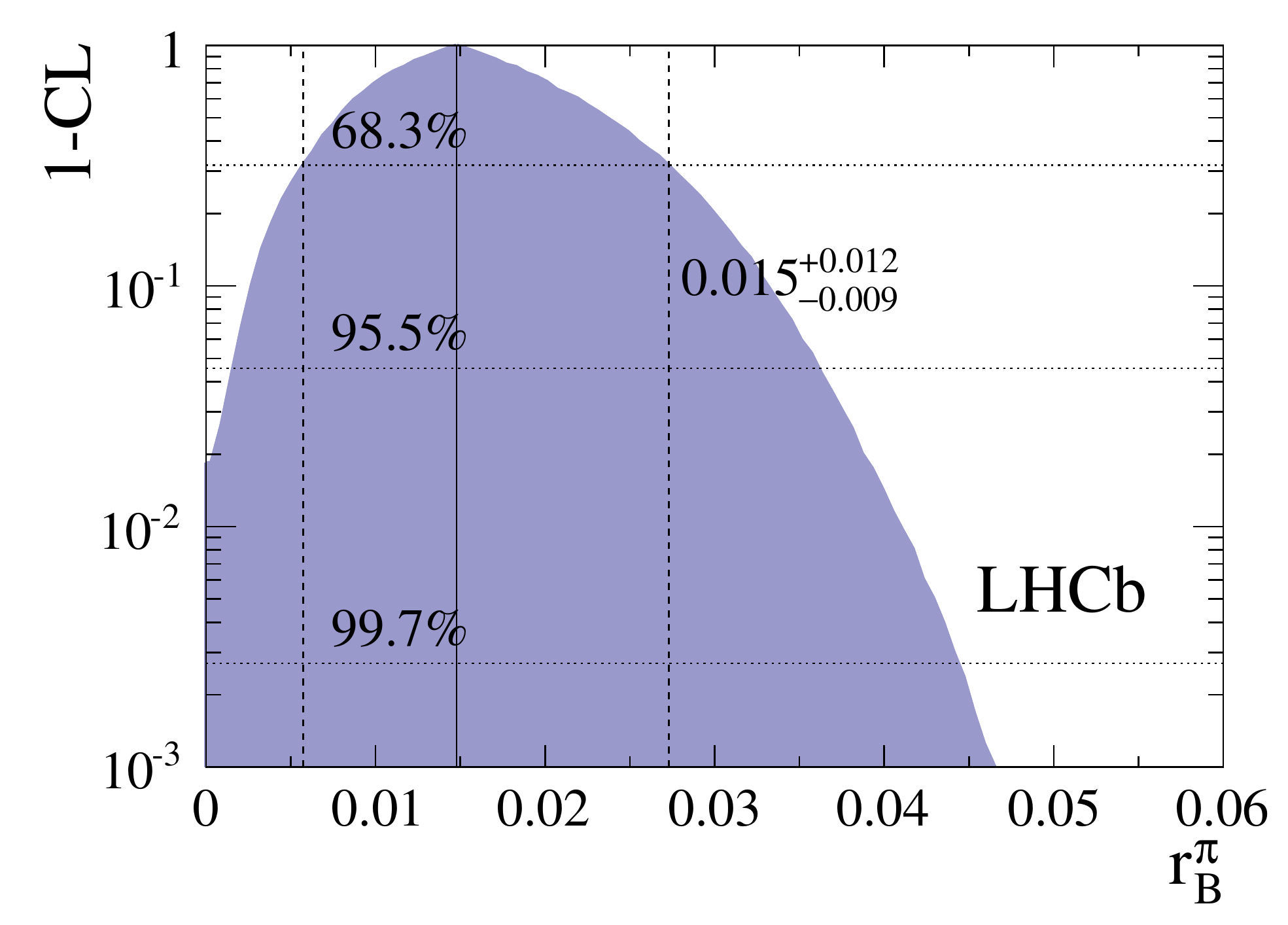}\put(190,135){d)}\end{overpic}\\
\begin{overpic}[width=.49\textwidth]{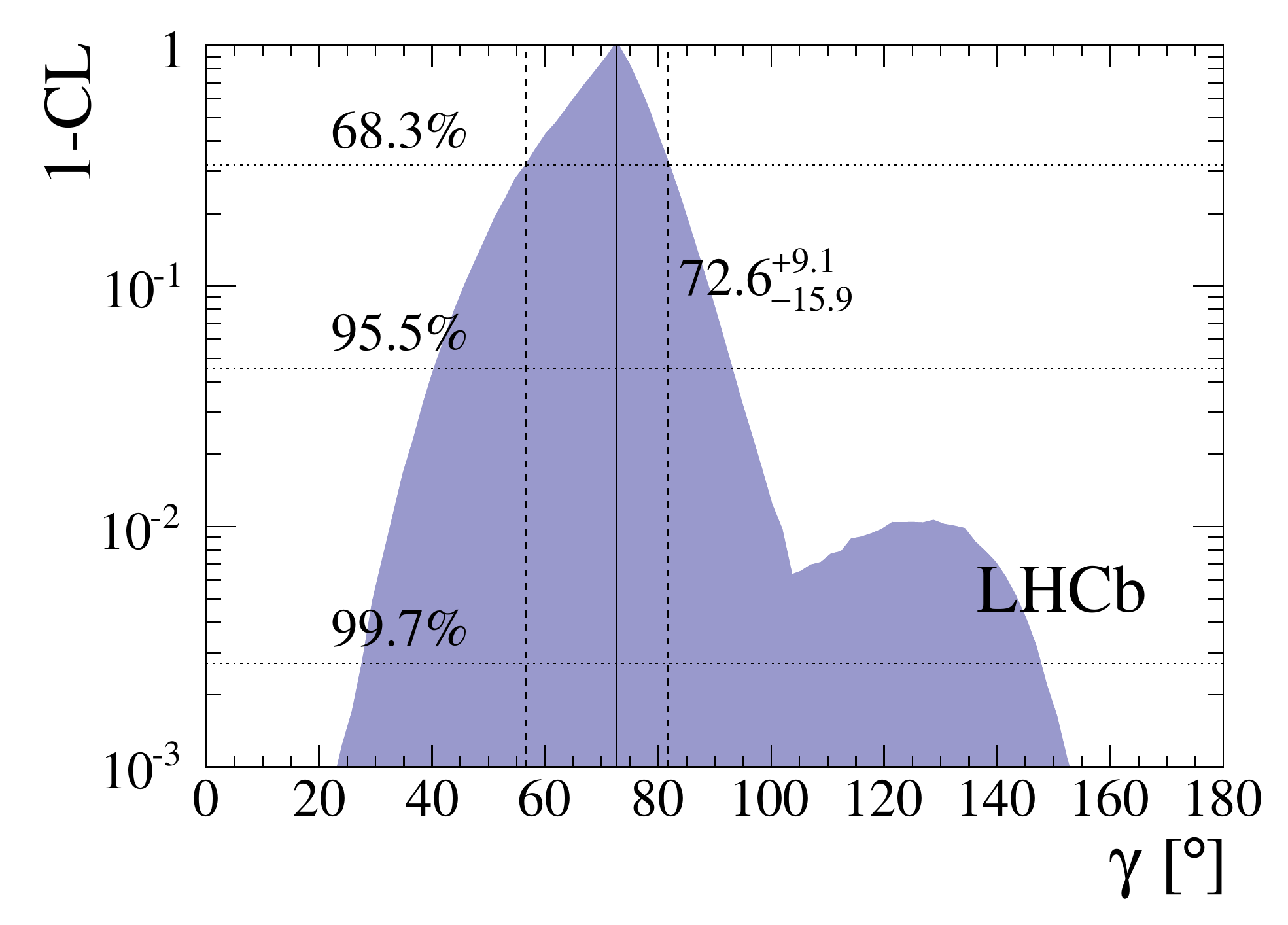}    \put(190,135){e)}\end{overpic}
\caption{\small Graphs showing \omcl for (a) \db, (b) \dbpi, (c) \rb, 
(d) \rbpi, and (e) \g, for the full \DzK and \Dzpi combination.
The reported numbers correspond to the best-fit values and the uncertainties
are computed using appropriate $68.3\%$ CL confidence intervals shown
in Table~\ref{tab:dkdpi}.}
\label{fig:dkdpi1log}
\end{figure}

\begin{figure}[!htb]
\centering
\begin{overpic}[width=.49\textwidth]{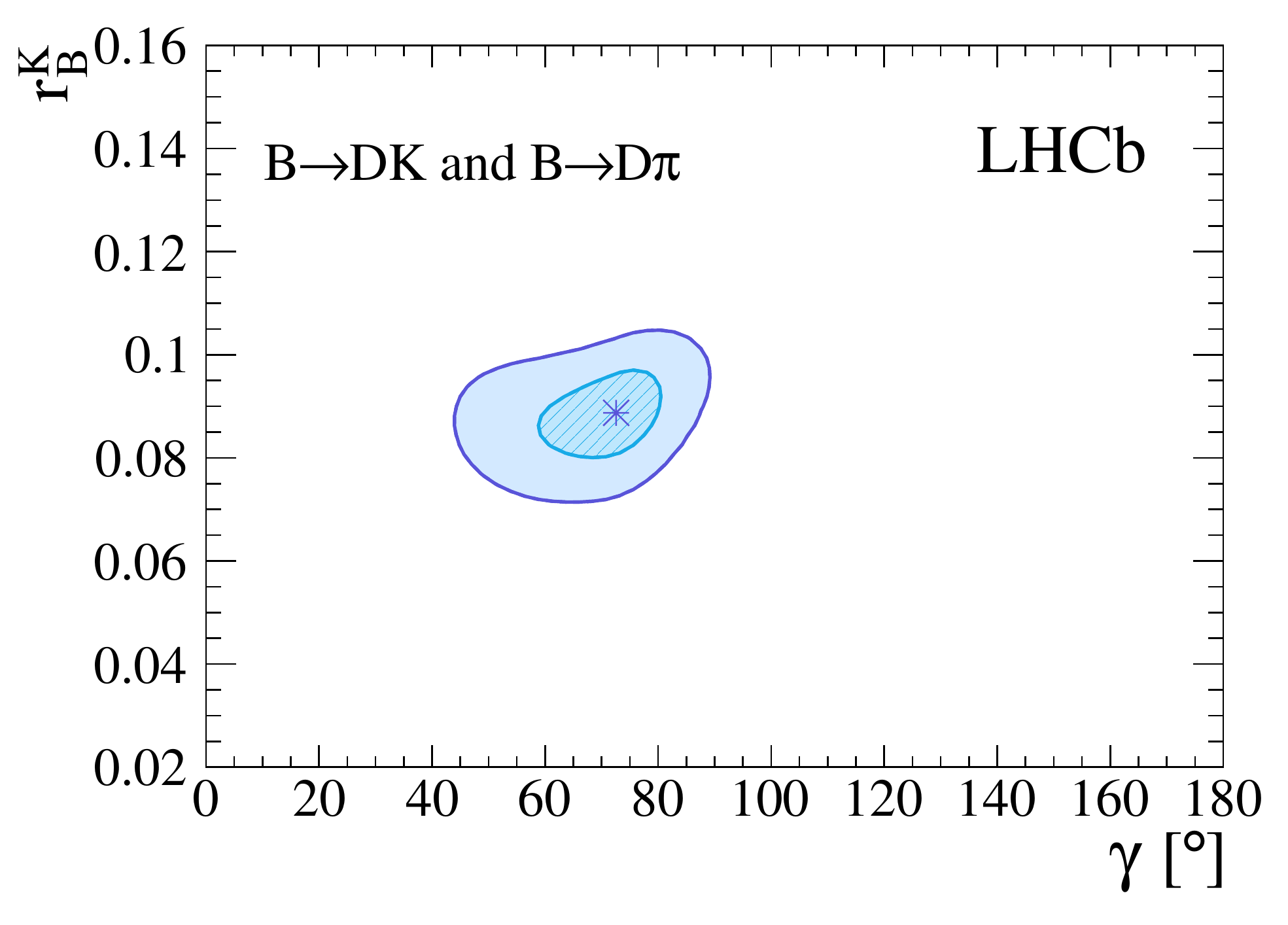} \put(190,40){a)}\end{overpic}
\begin{overpic}[width=.49\textwidth]{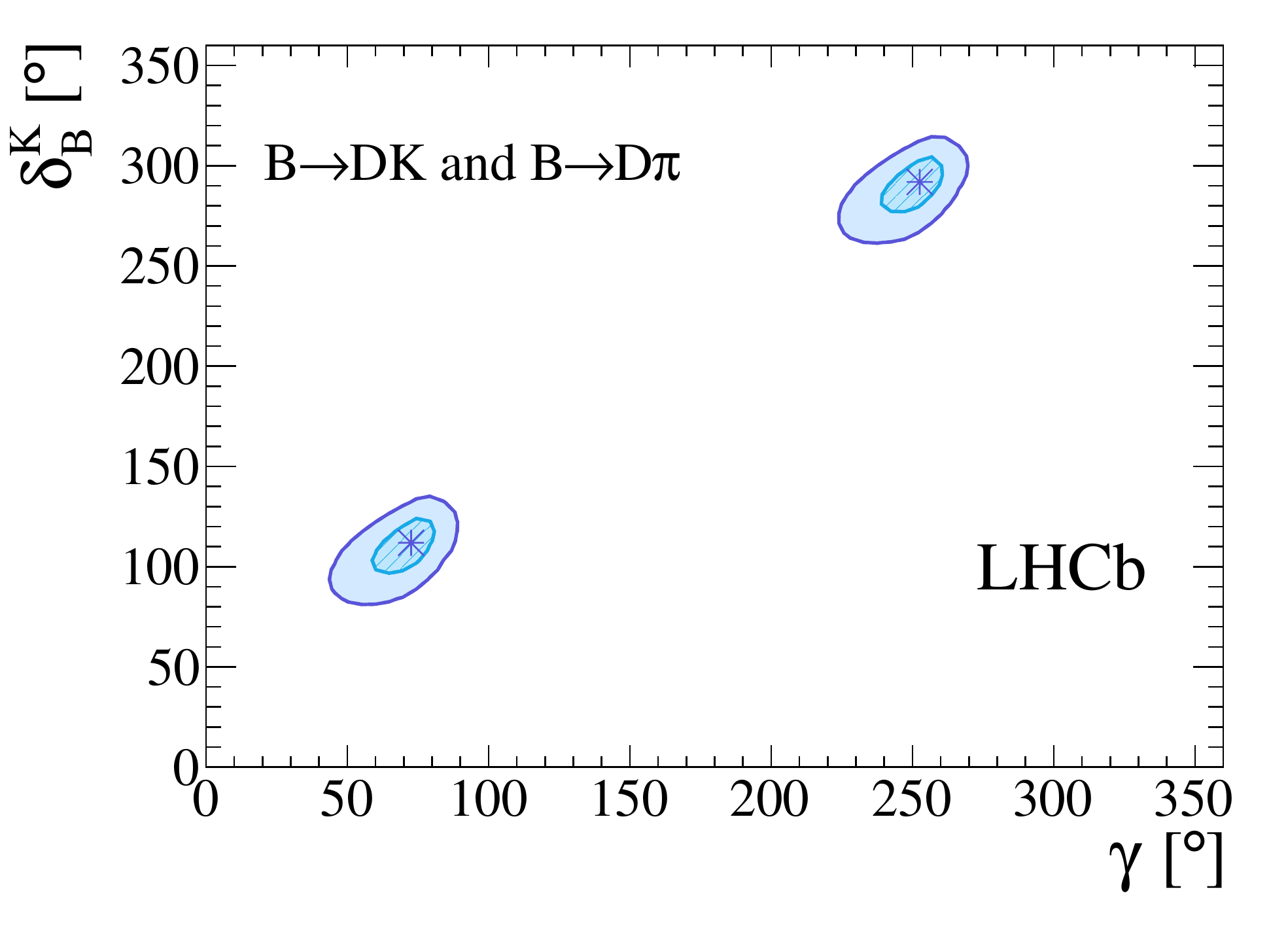} \put(190,40){b)}\end{overpic}\\
\begin{overpic}[width=.49\textwidth]{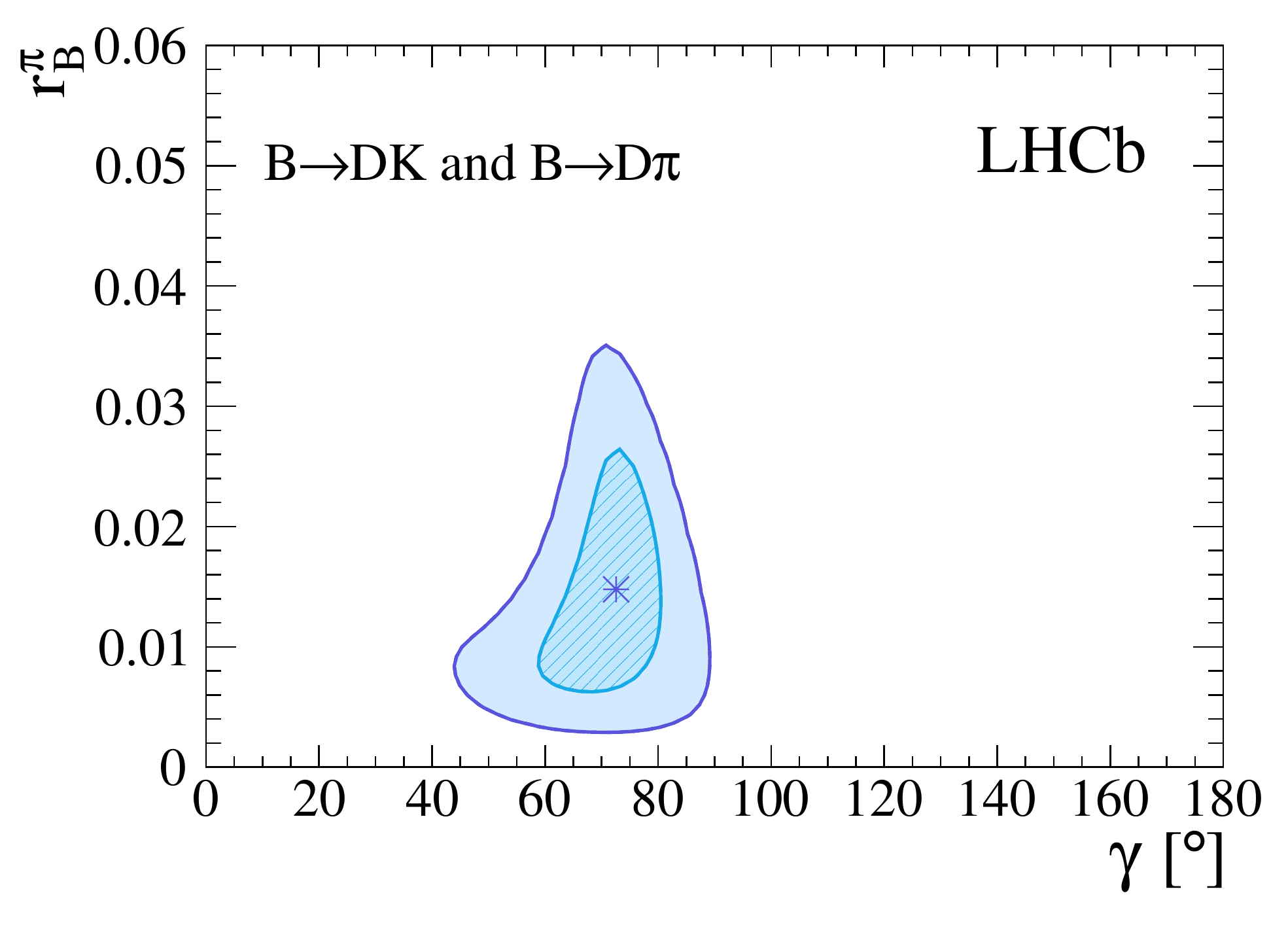}\put(190,40){c)}\end{overpic}
\begin{overpic}[width=.49\textwidth]{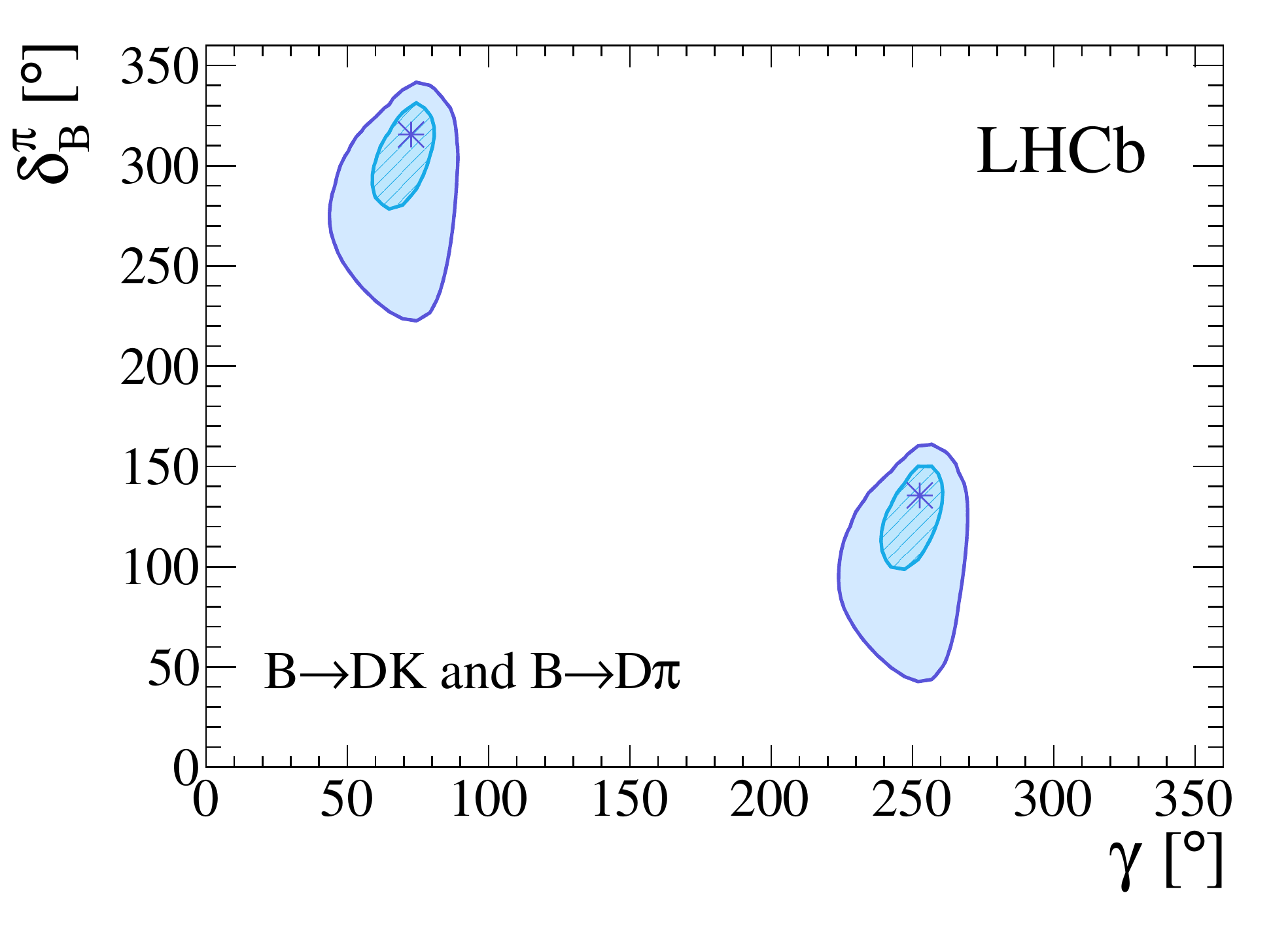}\put(190,40){d)}\end{overpic}\\
\caption{\small Profile likelihood contours of 
(a) \g vs.~\rb, 
(b) \g vs.~\db, 
(c) \g vs.~\rbpi, 
and (d) \g vs.~\dbpi, for the full
\DzK and \Dzpi combination.
The contours are the $n\sigma$ 
profile likelihood contours, where $\Delta\chi^2=n^2$ with $n=1,2$.
The markers denote the best-fit values.
Subfigures (b) and (d) show the full angular range to visualize the
symmetry, while subfigures (a) and (c) are expressed modulo $180^{\circ}$.
}
\label{fig:2dplots1}
\end{figure}

Three different combinations are presented.
First, only the parts corresponding to \BpmDK decays
of the two- and four-body GLW/ADS 
measurements~\cite{Aaij:2012kz,Aaij:2013aa}
are combined with the GGSZ~\cite{Aaij:2012hu} measurement. Then, only the \BpmDpi
parts of the two- and four-body GLW/ADS measurements are combined.
Finally, the full \BpmDh combination is computed.
It is difficult to disentangle the \BpmDK and \BpmDpi
measurements, because the observed ratios of Eq.~\ref{eq:rkpi}
necessarily contain information on both systems. 
These ratios are therefore included in the \BpmDK
combination, but not in the \BpmDpi combination. 
To include them in the \BpmDK combination, the
denominator in the second term of Eq.~\ref{eq:rkpi1} is assumed
to equal unity, neglecting a correction smaller than $0.04$, such
that effects of hadronic parameters in the \BpmDpi system are avoided.
The separate \DzK (\Dzpi) combination contains 29 (22) observables,
and the full combination contains 38 observables,
as 13 observables from CLEO, HFAG, and Ref.~\cite{LHCb_Dmixing} are common to both
separate combinations.
The results are summarised in Tables~\ref{tab:dk}--\ref{tab:dkdpi},
and illustrated in Figs.~\ref{fig:dk1log}--\ref{fig:dkdpi1log}.
The equations of Sect.~\ref{sec:inputs} are invariant under
the simultaneous transformation $\g\to\g+180^{\circ}$, $\delta\to\delta+180^{\circ}$,
where $\delta=\db, \dbpi$. 
All results on \g, \db, and
\dbpi are expressed modulo $180^{\circ}$, and only the solution
most consistent with the average computed by CKMfitter and UTfit is shown.
Figure~\ref{fig:2dplots1} shows two-dimensional profile likelihood contours
of the full combination, where the discrete symmetry is apparent in subfigures (b)
and (d).
The \DzK combination results in confidence intervals for \g 
that are symmetric and almost Gaussian up to $95\%$ CL.
Beyond that a secondary, local minimum of $\chi^2(\vec{\alpha}'_{\rm min})$
causes a much enlarged interval at $99\%$ CL.
The \Dzpi combination results in unexpectedly small confidence intervals
at $68\%$ CL. This can be explained by an upward fluctuation of 
\rbpi, since again the uncertainty of \g scales roughly like $1/\rbpi$.
The ratio \rbpi is expected to be 
$\rbpi\approx |(V_{ub}^* V_{cd}^{\phantom{*}})/(V_{cb}^* V_{ud}^{\phantom{*}})| \times |C|/|T+C| \approx 0.006$,
where $C$ and $T$ describe the magnitudes of the colour-suppressed
and tree amplitudes governing \BpmDpi decays, with their numerical values 
estimated from Ref.~\cite{Fleischer:2011uq}.
Within the $95\%$ CL interval, \rbpi is well consistent with
this expectation, and no constraints on \g are set.
The high value of \rbpi also affects the full combination.

\begin{table}[!h]
\centering
\caption{\small Confidence intervals and best-fit values of the \DzK 
combination for \g, \db, and \rb.}
\label{tab:dk}
\begin{tabular}{lc}
\hline
Quantity & \DzK combination \\
\hline
\g      & $\gDKCentral^\circ$ \\
68\% CL & $\gDKOnesig^\circ$ \\
95\% CL & $\gDKTwosig^\circ$ \\
\hline\\[-2.5ex]
\db     & $\dbDKCentral^\circ$ \\
68\% CL & $\dbDKOnesig^\circ$ \\
95\% CL & $\dbDKTwosig^\circ$ \\
\hline\\[-2.5ex]
\rb     & \rbDKCentral \\
68\% CL & $\rbDKOnesig$ \\
95\% CL & $\rbDKTwosig$ \\
\hline
\end{tabular}
\end{table}

\begin{table}[!h]
\centering
\caption{\small Confidence intervals and best-fit values for the \Dzpi combination
for \g, \dbpi, and \rbpi.
The corrections to the \g intervals for undercoverage and neglected systematic
correlations, as described in Sect.~\ref{sec:validation}, are not yet applied.}
\label{tab:dpi}
\begin{tabular}{lc}
\hline
Quantity & \Dzpi combination \\
\hline
\g       & $\gDpiCentralI^\circ$ \\
68\% CL  & $\gDpiOnesigI^\circ\,\cup\,\gDpiOnesigII^\circ$ \\
95\% CL  & no constraint \\
\hline\\[-2.5ex]
\dbpi   & $\dbDpiCentralI^\circ$ \\
68\% CL & $\dbDpiOnesigII^\circ\,\cup\,\dbDpiOnesigI^\circ$ \\
95\% CL & no constraint  \\
\hline\\[-2.5ex]
\rbpi   & \rbDpiCentral \\
68\% CL & \rbDpiOnesig  \\
95\% CL & \rbDpiTwosig  \\
\hline
\end{tabular}
\end{table}

\begin{table}[!h]
\centering
\caption{\small Confidence intervals and best-fit values
for the \DzK and \Dzpi combination for
\g, \rb, \db, \rbpi, and \dbpi.
The corrections to the \g intervals for undercoverage and neglected systematic
correlations, as described in Sect.~\ref{sec:validation}, are not yet applied.}
\label{tab:dkdpi}
\begin{tabular}{lc}
\hline
Quantity & \DzK and \Dzpi combination \\
\hline
\g       & $\gDKDpiCentral^\circ$ \\
68\% CL  & $\gDKDpiOnesig^\circ$  \\
95\% CL  & $\gDKDpiTwosig^\circ$  \\
\hline\\[-2.5ex]
\rb      & \rbDKDpiCentral \\
68\% CL  & \rbDKDpiOnesig  \\
95\% CL  & \rbDKDpiTwosig  \\
\hline\\[-2.5ex]
\db      & $\dbDKDpiCentral^\circ$ \\
68\% CL  & $\dbDKDpiOnesig^\circ$    \\
95\% CL  & $\dbDKDpiTwosig^\circ$    \\
\hline
\rbpi    & \rbpiDKDpiCentral \\
68\% CL  & \rbpiDKDpiOnesig  \\
95\% CL  & \rbpiDKDpiTwosig  \\
\hline
\dbpi    & $\dbpiDKDpiCentral^\circ$ \\
68\% CL  & $\dbpiDKDpiOnesig$ \\
95\% CL  & no constraint \\
\hline
\end{tabular}
\end{table}

\clearpage

\section{Validation of results and systematic uncertainties}
\label{sec:validation}

To assess the agreement between the various input measurements, the probability $P$,
that the observed dataset agrees better with the best-fit model than a dataset
generated assuming that model, is considered.
It is computed in two different ways. 
A first estimation of $P$ is obtained as the $p$-value of a $\chi^2$ 
test on the value $\chi^2(\vec{\alpha}_{\rm min})$, assuming it follows the 
$\chi^2$ distribution with a number of degrees of freedom given by the 
difference of the the number of observables $n_{\rm obs}$ and the number 
of fit parameters $n_{\rm fit}$.
A more accurate approach is to generate
pseudodatasets $j$ at the best-fit value, and fit these datasets with all parameters
free. Then $P$ is given as the fraction of pseudoexperiments that satisfy
$(\chi^2_j > \chi^2_{\rm data})$.
For this test, the pseudoexperiments used for the plug-in method
are re-evaluated. 
The fit probability based on the $\chi^2$ distribution is
well consistent with that based on the pseudoexperiments,
as shown in Table~\ref{tab:gof}.

The statistical coverage of the plug-in method
is not guaranteed. Therefore the coverage is computed
at the best-fit point for each of the three combinations.
This is done by generating pseudodatasets at the best-fit
point, and then, for each dataset, computing
the $p$-value of the best-fit point using the plug-in
method. The coverage is then defined as the fraction 
$\alpha$ in which the  
best-fit value of $\gamma$ has a larger $p$-value than
$\eta=68.27\%$, $\eta=95.45\%$, and 
$\eta=99.73\%$, for 1-, 2-, $3\sigma$, respectively.
The plug-in method slightly undercovers in the \Dzpi and full combinations, as shown in 
Table~\ref{tab:coverage}. The DK combination has exact coverage.
The same table also contains
the coverage of the simpler interval setting approach, in which the confidence
intervals are defined by $\Delta\chi^2=n^2$, where
$n=1,2,3$. The profile likelihood approach was found to significantly
undercover. The \DzK combination has exact coverage. For the \Dzpi and
full combinations, the final plug-in confidence intervals
(Tabs.~\ref{tab:dpi},~\ref{tab:dkdpi})
are scaled up by 
factors $\eta/\alpha$, taken from Table~\ref{tab:coverage}.

In addition the confidence intervals were cross-checked using
a method inspired by Berger and Boos~\cite{bergerboos}. Instead
of setting the nuisance parameters $\vec{\theta}$ to their best-fit 
values when
computing the $p$-value, $p(\gamma_0,\theta)$, 
$n_{\rm BB}=50$ alternative points are chosen, drawn
from an $(n_{\rm fit}-1)$-dimensional uniform distribution over
a restricted region $C_\beta$. Then, the $p$-value is given as
$p_{\rm BB} = \max_{\vec{\theta}\in C_{\beta}} p(\gamma_0,\vec{\theta}) + \beta$.
Here, $\beta$ is the probability that $\vec{\theta}$ lies outside $C_\beta$, and 
$C_\beta$ is chosen large enough such that $\beta<10^{-4}$. This method
is more conservative than the nominal plug-in method, but is guaranteed
to not undercover for $n_{\rm BB}\to\infty$. The resulting intervals are
only slightly larger than the nominal ones.

\begin{table}[!b]
\centering
\caption{\small Numbers of observables $n_{\rm obs}$, numbers of free
parameters in the fit $n_{\rm fit}$, the minimum $\chi^2$ at the
best-fit point, and fit probabilities of the best-fit point for the
three combinations. The quoted uncertainties are due to the limited
number of pseudoexperiments.}
\label{tab:gof}
\begin{tabular}{lccccc}
\hline\\[-2.5ex]
Combination & $n_{\rm obs}$ & $n_{\rm fit}$ & $\chi^2_{\rm min}$ 
& $P$[\%] ($\chi^2$ distribution) & $P$[\%] (pseudoexperiments) \\
\hline\\[-2.5ex]
\DzK  & 29 & 15 & $10.48$ & $72.6$ & $73.9\pm0.2$ \\
\Dzpi & 22 & 14 & $6.28$  & $61.6$ & $61.2\pm0.3$ \\
full  & 38 & 17 & $13.06$ & $90.6$ & $90.9\pm0.1$ \\
\hline                                                    
\end{tabular}                                    
\end{table}

\begin{table}[htb!]
\centering
\caption{\small
Coverage fraction $f_{\rm in} = N_{\rm in}/N$ for $\gamma$ at its best measured 
value for 1-, 2-, and $3\sigma$ intervals, for the plug-in method and the simpler
approach based on the profile likelihood.
The quoted uncertainties are due to the limited number of pseudoexperiments.}
\label{tab:coverage}
\begin{tabular}{cccc}  
\hline
Combination    & $\eta$   & $\alpha$ (plug-in)    & $\alpha$ (profile likelihood) \\
\hline
\DzK           & $0.6827$ $(1\sigma)$ & $0.6874 \pm 0.0050$ & $0.6508 \pm 0.0051$ \\
               & $0.9545$ $(2\sigma)$ & $0.9543 \pm 0.0023$ & $0.9414 \pm 0.0025$ \\
               & $0.9973$ $(3\sigma)$ & $0.9952 \pm 0.0007$ & $0.9947 \pm 0.0008$ \\
\hline
\Dzpi          & $0.6827$ $(1\sigma)$ & $0.5945 \pm 0.0053$ & $0.5105 \pm 0.0054$ \\
               & $0.9545$ $(2\sigma)$ & $0.9391 \pm 0.0026$ & $0.9238 \pm 0.0029$ \\
               & $0.9973$ $(3\sigma)$ & $0.9960 \pm 0.0007$ & $0.9919 \pm 0.0010$ \\
\hline
\DzK and \Dzpi & $0.6827$ $(1\sigma)$ & $0.6394 \pm 0.0050$ & $0.5839 \pm 0.0051$ \\
               & $0.9545$ $(2\sigma)$ & $0.9374 \pm 0.0025$ & $0.9112 \pm 0.0030$ \\
               & $0.9973$ $(3\sigma)$ & $0.9929 \pm 0.0009$ & $0.9912 \pm 0.0010$ \\
\hline
\end{tabular}
\end{table}

For the two-body and four-body GLW/ADS analyses no information
on systematic correlations is available. Consequently, they are
assumed to be zero in the nominal combinations. Their 
possible influence is assessed by computing the effect of a large number of
random correlation matrices on the expected confidence intervals.
A maximum correlation of $75\%$ is considered in the random matrices.
The expected intervals are computed by generating pseudodatasets
at the best-fit points of the three combinations, and then, for each
pseudodataset, by computing its profile $\Delta\chi^2$ curve, and taking
the average of these curves. The \DzK combination is unaffected.
The \Dzpi combination, however, is affected to a large extent, as the values
of several observables are limited by systematic uncertainties.
Conservatively, the maximum of the $p$-values observed for all 
random correlation matrices is considered. The nominal $1\sigma$ intervals are 
asymmetrically enlarged by $12\%$ to match the maximum.
The full combination is only slightly affected. The systematic uncertainty
is fully concentrated in the lower side of the interval. Therefore, a systematic
uncertainty of $2.5^\circ$ $(5.0^\circ)$ is added in quadrature to the lower
$1\sigma$ $(2\sigma)$ errors.

The linearity of the combination procedure was checked by computing
values for all observables using the best-fit point of the full combination
and the relations from Section~\ref{sec:inputs}. Assuming the experimental covariances,
the best-fit point was perfectly reproduced, and the procedure was found to
be unbiased.

In summary, the \DzK combination does not require corrections.
In case of the \Dzpi and full combinations, the intervals are enlarged to account
for both neglected systematic correlations and undercoverage.

\section{Conclusion}
\label{sec:conclusion}

A combination of recent \lhcb results~\cite{Aaij:2012kz,Aaij:2012hu,Aaij:2013aa}
is used to measure the CKM angle \g.
The decays \BpmDK and \BpmDpi are used, where the $D$ meson decays into $\Kp\Km$, 
$\pip\pim$, $K^\pm \pi^\mp$, $\KS\pip\pim$, $\KS \Kp\Km$, or $K^\pm \pi^\mp \pi^+ \pi^\mp$ final states. 
The effect of \Dz--\Dzb mixing is taken into account in the ADS analysis of both
\BpmDK and \BpmDpi decays.
Using only \BpmDK results, a best-fit value in $[0,180]^\circ$ of
$\g = \gDKCentral^\circ$ is found and 
confidence intervals are set using a frequentist procedure
\begin{align*}
\g \in \gDKOnesigC^\circ \quad &{\rm at~68\%\,CL}\,,\\
\g \in \gDKTwosigC^\circ \quad &{\rm at~95\%\,CL}\,.
\end{align*}
Taking the best-fit value as central value, 
the first interval is translated to
\begin{equation*}
\g = (\gDKCentralPMC)^{\circ}\,{\rm at~68\%\,CL}\,.
\end{equation*}
At $99\%$ CL a second (local) minimum contributes
to the interval.
When combining results from \BpmDpi decays alone, 
a best-fit value of $\g = \gDpiCentralI^\circ$ is found and the
following confidence intervals are set
\begin{align*}
\g \in 
\gDpiOnesigIC^\circ \quad \cup \quad
\gDpiOnesigIIC^\circ \quad &{\rm at~68\%\,CL}\,,
\end{align*}
and no constraint is set at $95\%$ CL.
For the first time, information from \BpmDpi decays 
is included in a combination. 
When these results are included, the best-fit value becomes $\g = \gDKDpiCentral^\circ$ 
and the following confidence intervals are set
\begin{align*}
\g \in \gDKDpiOnesigC^\circ   \quad &{\rm at~68\%\,CL}\,,\\
\g \in \gDKDpiTwosigC^\circ   \quad &{\rm at~95\%\,CL}\,.
\end{align*}
All quoted values are modulo $180^\circ$.
The coverage of our frequentist method was evaluated and found
to be exact when combining \BpmDK results alone, and accurate within 4\% (2\%) at $1\sigma$
($2\sigma$) when combining \BpmDK and \BpmDpi results. The
final intervals have been scaled up to account for this undercoverage,
and to account for neglected systematic correlations.

\section*{Acknowledgements}

\noindent We express our gratitude to our colleagues in the CERN
accelerator departments for the excellent performance of the LHC. We
thank the technical and administrative staff at the LHCb
institutes. We acknowledge support from CERN and from the national
agencies: CAPES, CNPq, FAPERJ and FINEP (Brazil); NSFC (China);
CNRS/IN2P3 and Region Auvergne (France); BMBF, DFG, HGF and MPG
(Germany); SFI (Ireland); INFN (Italy); FOM and NWO (The Netherlands);
SCSR (Poland); ANCS/IFA (Romania); MinES, Rosatom, RFBR and NRC
``Kurchatov Institute'' (Russia); MinECo, XuntaGal and GENCAT (Spain);
SNSF and SER (Switzerland); NAS Ukraine (Ukraine); STFC (United
Kingdom); NSF (USA). We also acknowledge the support received from the
ERC under FP7. The Tier1 computing centres are supported by IN2P3
(France), KIT and BMBF (Germany), INFN (Italy), NWO and SURF (The
Netherlands), PIC (Spain), GridPP (United Kingdom). We are thankful
for the computing resources put at our disposal by Yandex LLC
(Russia), as well as to the communities behind the multiple open
source software packages that we depend on.

\addcontentsline{toc}{section}{References}
\bibliographystyle{LHCb}
\bibliography{references}

\clearpage
\section{Supplemental material}
\label{sec:extramaterial}

The following material is to be made publicly available,
but is not to be included in the paper.

\begin{figure}[htb]
  \centering
  \includegraphics[width=.45\textwidth]{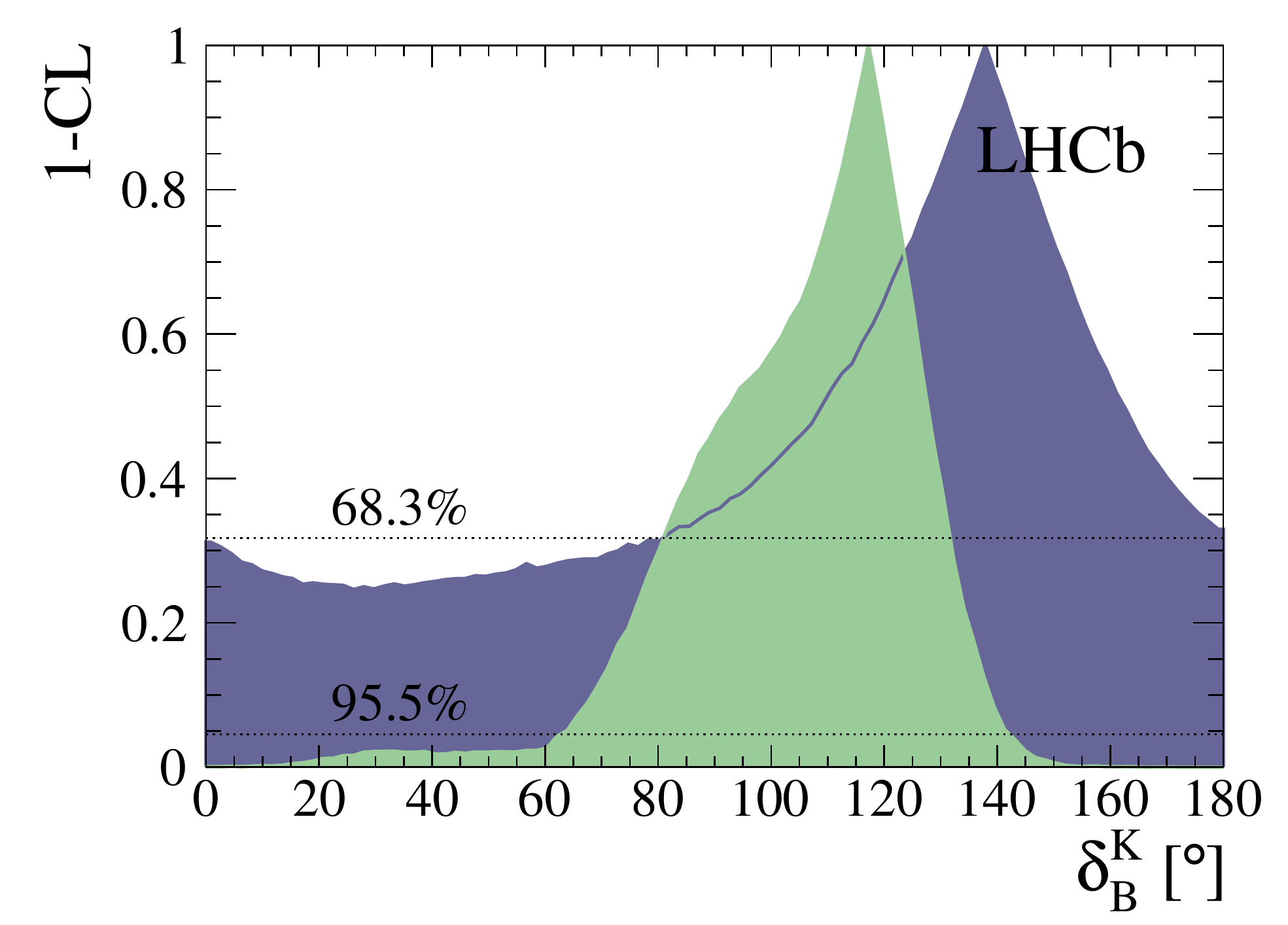}
  \includegraphics[width=.45\textwidth]{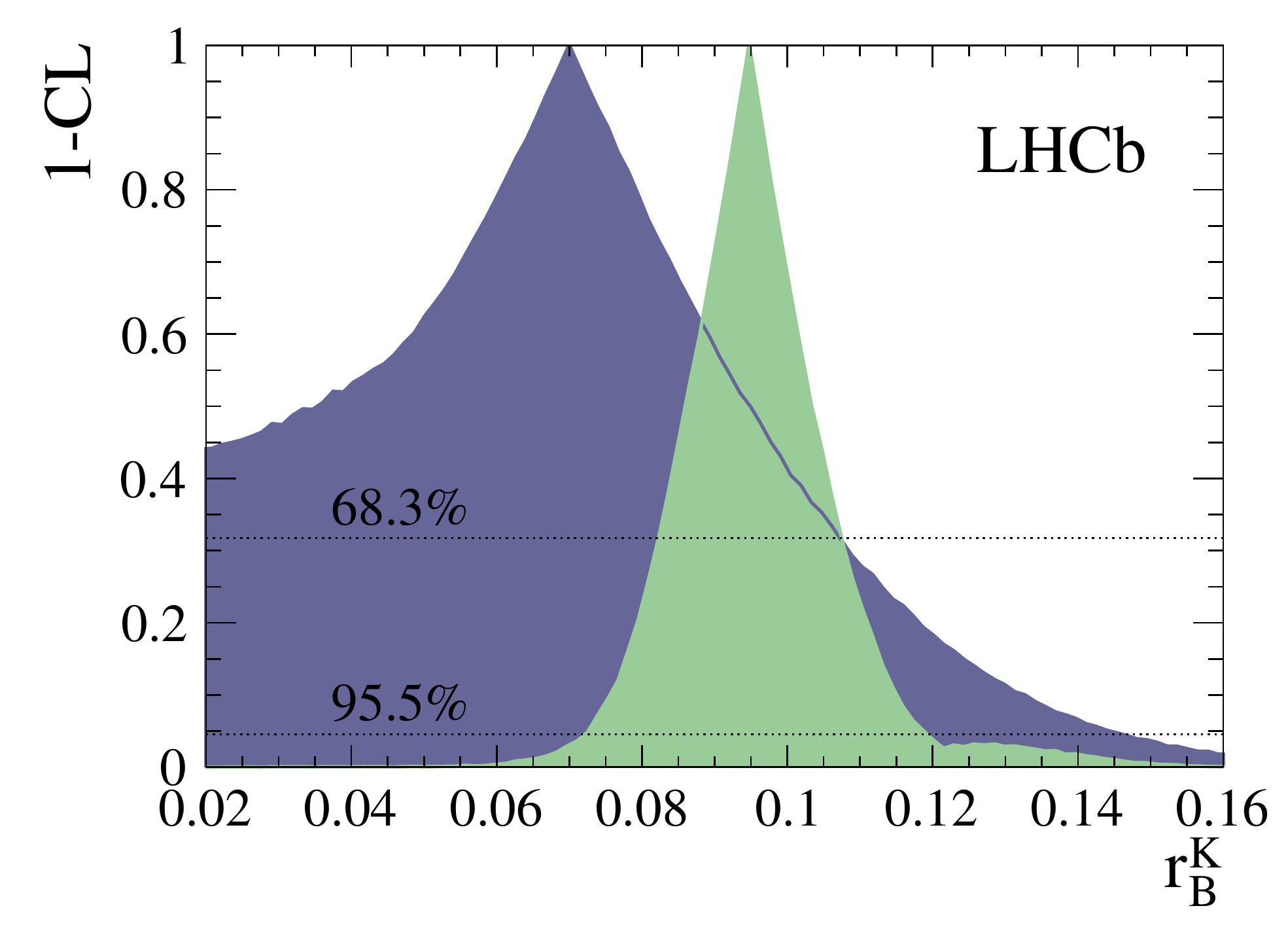}
  \includegraphics[width=.45\textwidth]{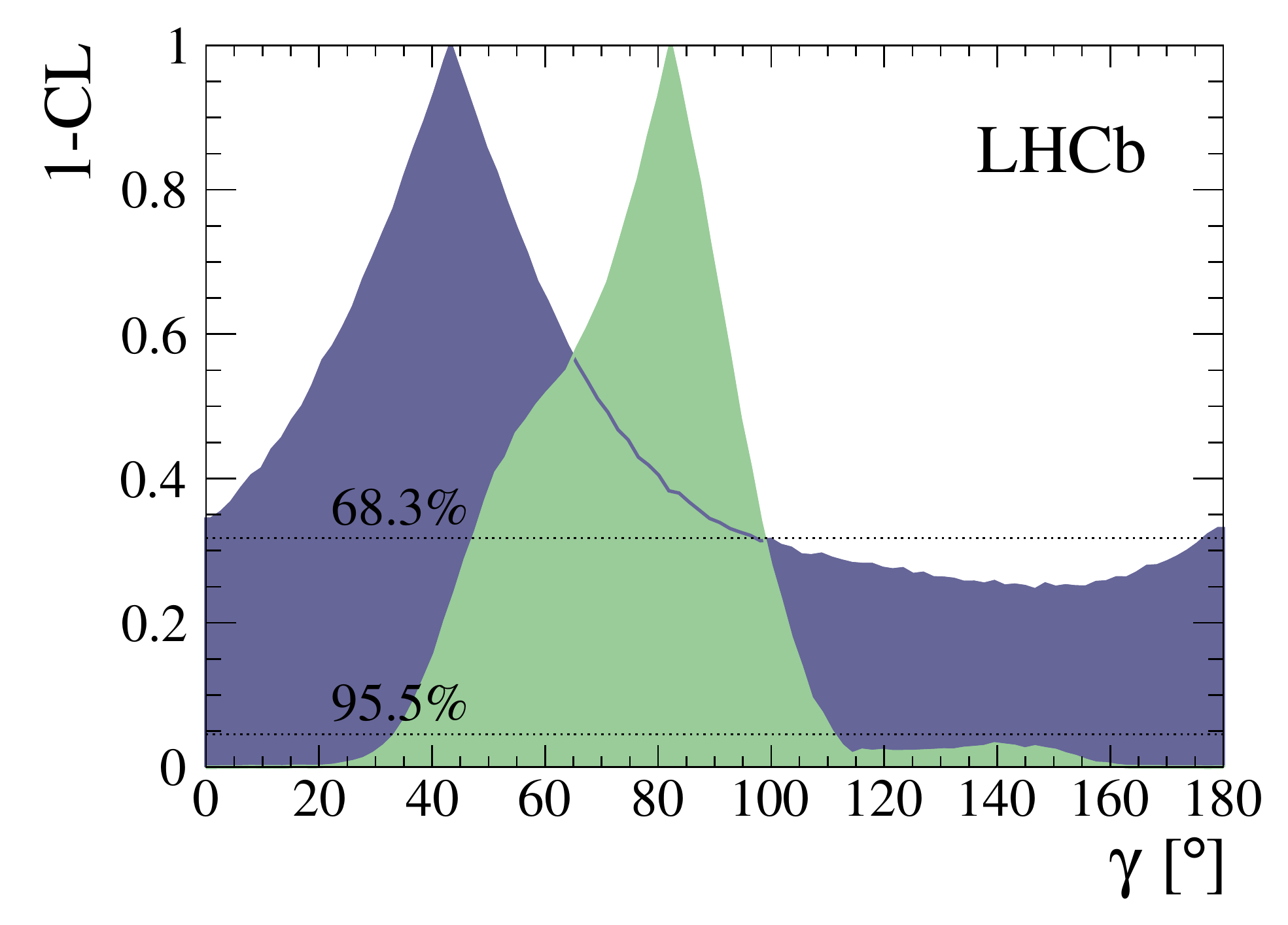}
  \caption{\small Graphs showing \omcl for \db, \rb, and \g, separately 
  for the GLW/ADS (light green) and GGSZ (dark purple) parts
  of the \DzK-only combination.}
  \label{fig:dzk_glwads_ggsz}
\end{figure}

\begin{figure}[htb]
  \centering
  \includegraphics[width=.45\textwidth]{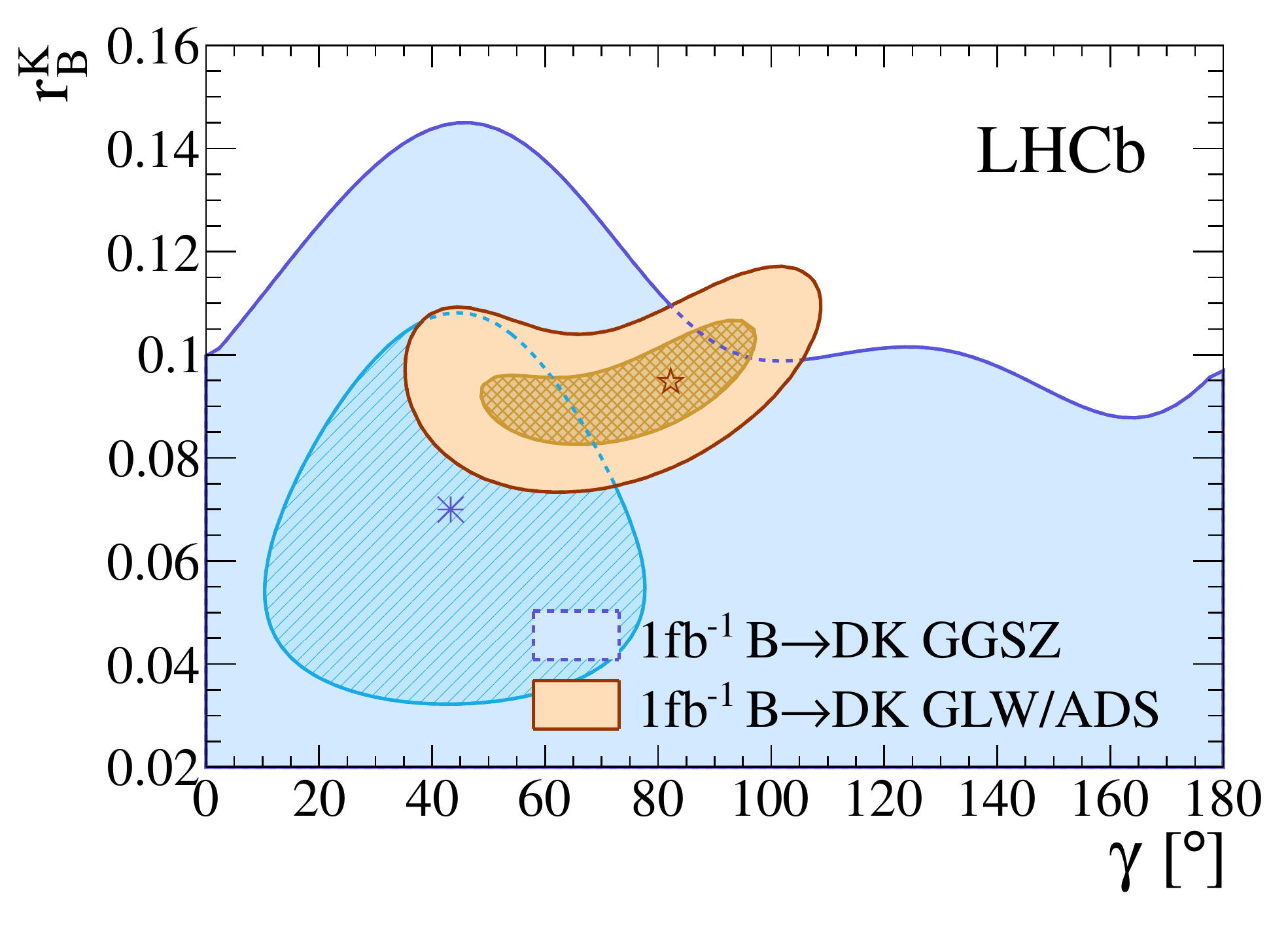}
  \includegraphics[width=.45\textwidth]{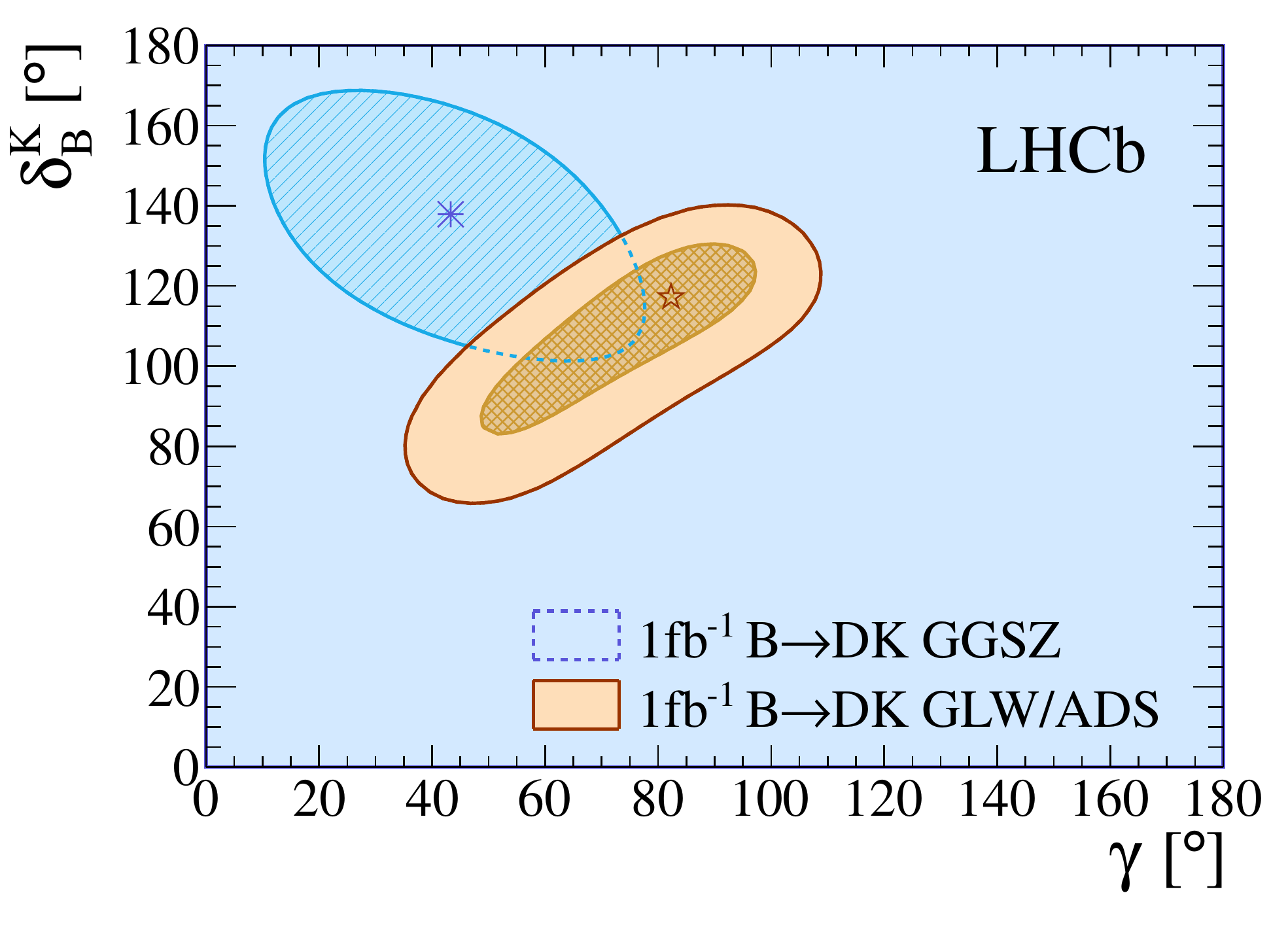}
  \includegraphics[width=.45\textwidth]{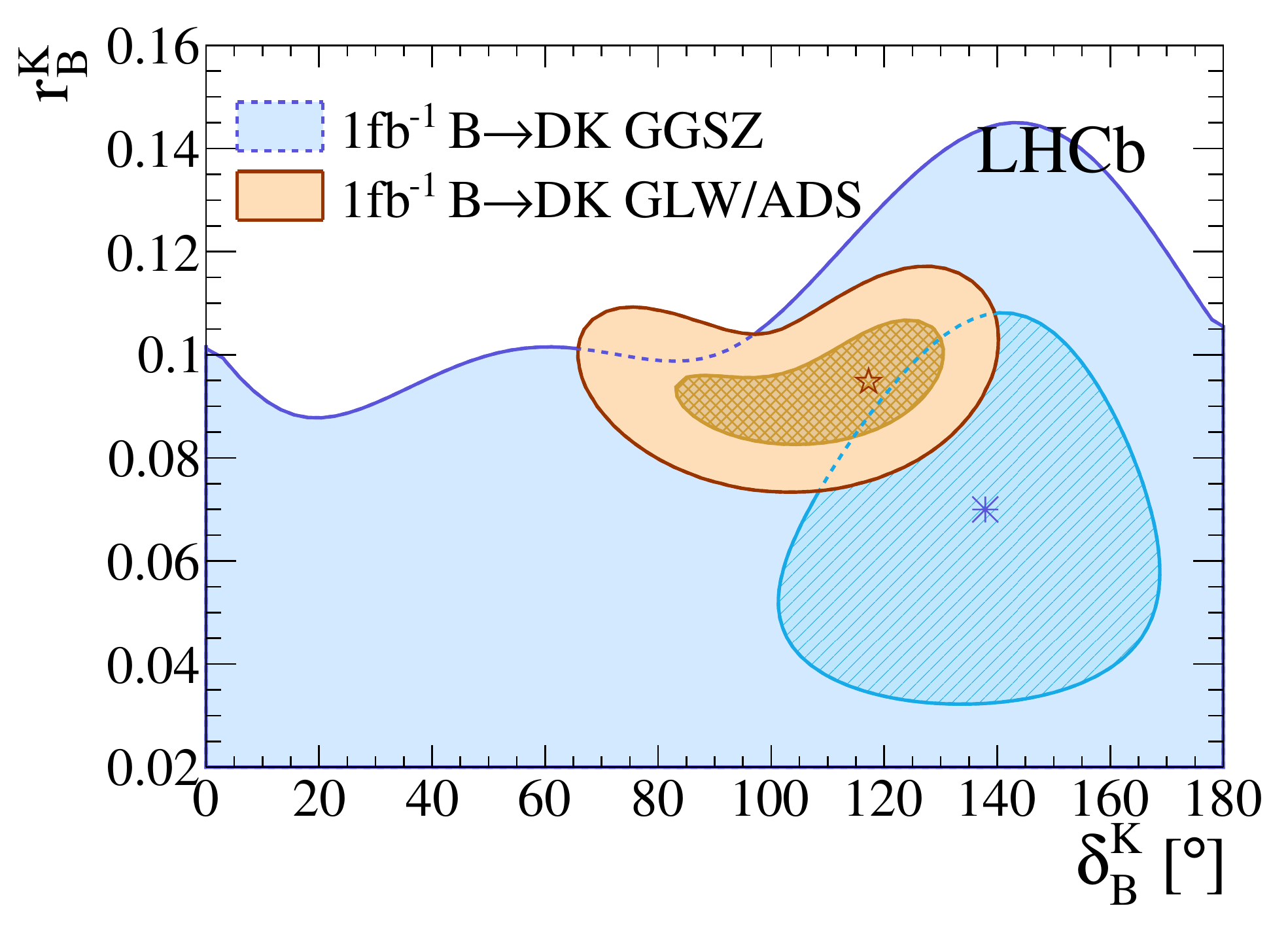}
  \caption{\small Profile likelihood contours, separately for the GGSZ (blue) and 
  two-body and four-body GLW/ADS (orange) parts of the \DzK only combination.
  The contours are the usual $n\sigma$ profile likelihood contours,
  where $\Delta\chi^2=n^2$ with $n=1,2$.
  The markers correspond to the best-fit points.}
  \label{fig:2dplots_dk_ggszglwads}
\end{figure}

\begin{figure}[htb]
  \centering
  \includegraphics[width=.45\textwidth]{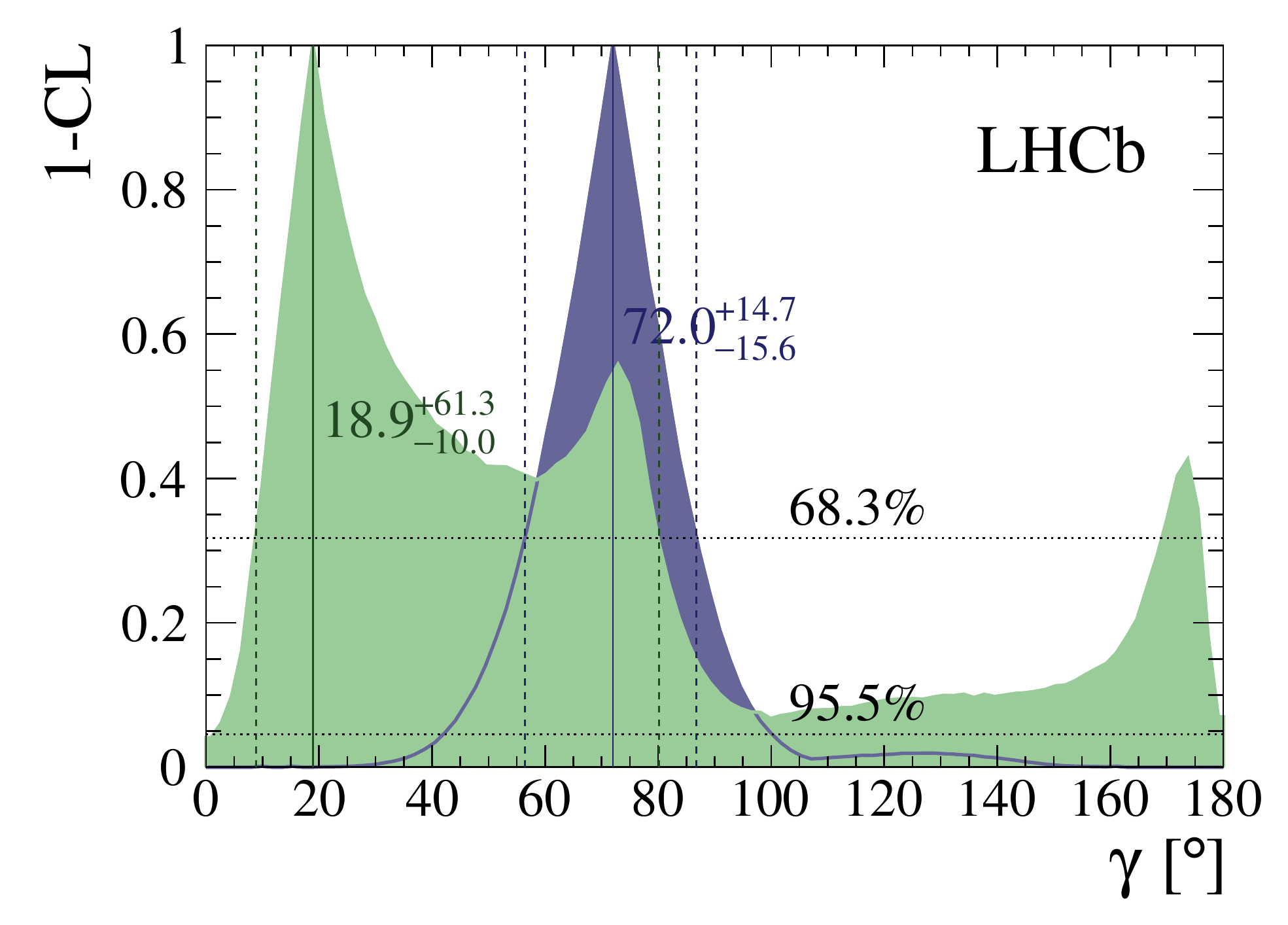}
  \caption{\small Graphs showing \omcl for \g, separately 
  for the \DzK-only combination (dark purple) and \Dzpi-only combination (light green).}
  \label{fig:dzk_dzpi}
\end{figure}

\end{document}